%% file: arxivafterR1_journal_caching_with_matrix_v1.tex
\documentclass[journal,12pt,draftclsnofoot,onecolumn,english]{IEEEtran}

\usepackage{epsfig}
\usepackage{times}
\usepackage{float}
\usepackage{afterpage}
\usepackage{sansmath}
\usepackage{amsmath}
\usepackage{amstext,cite}
\usepackage{amssymb,bm}
\usepackage{latexsym}
\usepackage{color}
\usepackage{graphicx}
\usepackage{amsmath}
\usepackage{amsthm}
\usepackage{graphicx}
\usepackage[center]{caption}
\usepackage{pstricks}
\usepackage{caption}
\usepackage{subcaption}
\usepackage{booktabs}
\usepackage{multicol}
\usepackage{lipsum}
\usepackage[T1]{fontenc}
\usepackage{hyperref}
\usepackage{aecompl}
\usepackage{mathrsfs}

\usepackage{arydshln}

\usepackage{soul}
\usepackage{cancel}

\usepackage{changes}
\definechangesauthor[color=red,name={Mingyue Ji}]{MJ}

\allowdisplaybreaks

\input{macros}

\newcommand{\dt}[1]{{\blue #1}}

\usepackage{amsmath}
\usepackage{tikz}
\usetikzlibrary{calc}

\usepackage[nodisplayskipstretch]{setspace} 

\begin{document}

\title{Cache-Aided Matrix Multiplication Retrieval} 
\author{
Kai~Wan,~\IEEEmembership{Member,~IEEE,} 
Hua~Sun,~\IEEEmembership{Member,~IEEE,}
Mingyue~Ji,~\IEEEmembership{Member,~IEEE,}  
Daniela Tuninetti,~\IEEEmembership{Senior~Member,~IEEE,}
and~Giuseppe Caire,~\IEEEmembership{Fellow,~IEEE}
\thanks{
A short version of this paper  has been accepted    in the 2021 IEEE 
International Symposium on Information Theory, Melbourne, Victoria, Australia.
} 
\thanks{
K.~Wan and G.~Caire are with the Electrical Engineering and Computer Science Department, Technische Universit\"at Berlin, 10587 Berlin, Germany (e-mail:  kai.wan@tu-berlin.de; caire@tu-berlin.de). The work of K.~Wan and G.~Caire was partially funded by the European Research Council under the ERC Advanced Grant N. 789190, CARENET.}
\thanks{
H.~Sun is with the Department of Electrical Engineering, University of North Texas, Denton, TX 76203, USA (email: hua.sun@unt.edu). The work of H. Sun was supported in part by NSF Award 2007108.
}
\thanks{
M.~Ji is with the Electrical and Computer Engineering Department, University of Utah, Salt Lake City, UT 84112, USA (e-mail: mingyue.ji@utah.edu). The work of M.~Ji was supported in part by NSF Awards 1817154 and 1824558.}
\thanks{
D.~Tuninetti is with the Electrical and Computer Engineering Department, University of Illinois Chicago, Chicago, IL 60607, USA (e-mail: danielat@uic.edu). The work of D.~Tuninetti was supported in part by NSF Award 1910309.}
}
\maketitle

\begin{abstract}
Coded caching is a promising technique to smooth out network traffic by storing part of the library content at the users’ local caches. The seminal work on coded caching for single file retrieval by Maddah-Ali and Niesen (MAN)  showed the existence of a global caching gain that scales with the total memory in the system, in addition to the known local caching gain in uncoded systems. This paper formulates a novel cache-aided matrix multiplication  retrieval  problem, 
 relevant  for data analytics and machine learning applications. 
In the considered problem, each cache-aided user requests the product of two matrices from the library. 
A structure-agnostic solution is to treat each possible matrix product as an independent file and use the MAN coded caching scheme for single file retrieval.  
This paper proposes two structure-aware schemes, which partition each matrix in the library by either rows or columns and let a subset of users cache some sub-matrices, that improve on the structure-agnostic scheme.
For the case where the library matrices are ``fat'' matrices, the structure-aware  row-partition scheme is shown to be order optimal under some constraint.
\end{abstract}


\section{Introduction}
\label{sec:intro}
It is predicted that an  order  of magnitude increase in network throughput is needed to support the tremendous growth of data traffic expected for the near future~\cite{cisco2014data}. 
Conventional technologies are severely limited towards the goal of achieving such a dramatic throughput gain. 
A clever usage of low-cost storage capacity on user devices to cache data plays a key role in the design of content distribution schemes.
Coded caching is an effective way to smooth out network traffic during peak traffic hours by jointly designing cache placement and coded delivery schemes. 
The coded caching strategy originally proposed by Maddah-Ali and Niesen (MAN) in~\cite{dvbt2fundamental} has the potential to trade off relatively cheap memory for expensive bandwidth, i.e., the total traffic load on the network is inversely proportional to the aggregate cache memory in the network, a phenomenon referred to as {\it global coded caching gain.}

The 
MAN original model consists of a server, with access to the whole library, that is connected to several cache-aided users through an error-free shared-link.
The MAN scheme contains two phases:
(i) {\it placement phase} (peak-off hours): each cache-aided user stores some bits in its local cache without knowledge of later demands;
(ii) {\it delivery phase} (peak-traffic hours): each user requests one file from the library and the server broadcasts coded packets to satisfy all users' requests simultaneously. The goal is to minimize the number of broadcasted bits for the worst-case demands, referred to as {\it worst-case load}.
It was surprisingly shown in~\cite{dvbt2fundamental} that if each bit in the library can be cached by $t$ users, the total load can potentially be reduced by $t+1$ times compared to the conventional uncoded caching scheme, in which the server  simply broadcasts to each user the uncached part of the demanded file.  
The MAN shared-link coded caching problem  for single file retrieval has been extended to a number of different network models (such as Device-to-Device networks~\cite{d2dcaching}, topological networks~\cite{hierarchicalcaching}, multi-server networks~\cite{multiserver}, wireless interference channels~\cite{interferencemanagement}, etc.) and different problems where reducing the communication cost is paramount (such as coded distributed computing~\cite{distributedcomputing}, coded data shuffling~\cite{speedup2018Lee,neartoptimalAttia2018,fundamentalshuffling2018,decentralizedDS2019wan}, etc.).

Matrix multiplication plays a key role in a wide variety of domains, such as for example data analytics, machine learning, and scientific computing~\cite{
speedup2018Lee,yu2020stragglermitigation,dutta2020optimalRT}.  
Recently, information theoretic coding techniques have been proposed for the distributed matrix multiplication problem~\cite{polyoptYu2017,yu2020stragglermitigation,dutta2020optimalRT,aliasgari2019private,chang2019privatesecure,jia2019matrixCSA,yu2020entangle}. 
In a distributed computing system 
a master node aims to compute the multiplication of two large-scale matrices with the help of workers, where the workers can only store and compute on small parts of the matrices.
Since workers may take different amounts of time to complete their assigned task, i.e., some are stragglers, the goal here is for the master node to recover the matrix product as soon as the number of responses received from the workers reaches the so-called recovery threshold.
Different coding schemes have been proposed 
to mitigate the impact of stragglers   on the completion time of a distributed computing task, such as polynomial codes~\cite{yu2020stragglermitigation,lagrange2019yu} and Matdot codes~\cite{dutta2020optimalRT}. 
Recently, distributed matrix multiplication for resilience against stragglers was extended to wireless channels~\cite{li2020wirelessmatrix}, where several users without local cache are connected to edge nodes with computation resources through a wireless link and where each user requests the product of a user-generated data matrix with a network-stored matrix. In this work, we are not interested in the problem of straggler mitigation, but rather in  the problem of reducing the communication load across a shared-link network.


This paper formulates a novel shared-link cache-aided matrix multiplication retrieval problem. 
With  the motivation   that matrix multiplication is one of the key building blocks
underlying many data analytics and machine learning algorithms,  we consider   the case where the cache-aided users request the multiplication of two matrices in the library, instead of a single file. 
For   example, each user aims to compute the linear correlation  between each two vectors of two vector sets,\footnote{\label{foot:linear correlation}
  Linear correlation is used to find the linear relationship between two numerically expressed variables, which has wide applications in lots of areas, such as 
engineering research (including pattern recognition~\cite{correlationpattern}, signal detection~\cite{signaldetection1990}, etc.)  
and
medical science~\cite{medicalcorrelation}.
} which can be seen as the multiplication of two matrices representing these two vector sets.

In our setting, the library contains $\Nsf$ files that are thought of as matrices of dimension $\ssf \times \rsf$ on some finite field. In the placement phase,  each of the $\Ksf$ users can store  up to $\Msf\ssf\rsf$ symbols from the library (corresponding to the size of up to $\Msf$ matrices). During the delivery phase, each user requests the product of two arbitrary matrices  in the library, which are not known in advance at the time of cache placement. 
Different from existing information theoretic distributed matrix multiplication works for straggler mitigation, {\bf we aim to apply coded caching strategies to the matrix multiplication retrieval problem with the goal of minimizing the load on the shared link between the server and the users  by leveraging the cached contents and performing coded multicast delivery.} 

\subsection{Main Contributions}
Our main contributions are as follows.
\begin{itemize}
\item
We formulate an information theoretic shared-link coded caching problem for matrix multiplication retrieval, where each user requests the product of two matrices in the library.
\item
We propose a structure-agnostic scheme that treats each possible demanded matrix product as an independent file and attains the load corresponding to the MAN coded caching problem for single file retrieval. 
\item
Then, we propose two coded caching schemes that leverage the specific structure of matrix multiplication. 
Different from the structure-agnostic matrix multiplication retrieval scheme, which lets the users directly cache 
 some entries  of the matrix products,  the proposed structure-aware schemes let each user cache 
 some entries of each matrix. One scheme partitions each library matrix into sub-matrices by rows and the other by columns. A subset of the users cache each sub-matrix, or some linear transformation of this sub-matrix. 
The delivery phase is designed so as to leverage the users' cached contents and the ``correlation'' among the elements of the demanded matrix products, i.e., the fact that some entries of a matrix product can be written as a function of the other entries of the same matrix product.
\item
When $\ssf \leq  \rsf$ (i.e., the library matrices are ``fat'' matrices), we prove that  the proposed row-partition scheme is order optimal   within a factor of $2$ under the constraint of uncoded cache placement (i.e., each user directly copies some entries of the matrices in the library into its local cache) and $\Nsf \geq 2\Ksf$. This is accomplished by proposing a novel genie-aided converse bound.

\end{itemize}

\subsection{Paper Organization}
The rest of this paper is organized as follows.
Section~\ref{sec:pre} gives  some 
 results used later in the paper.
Section~\ref{sec:model} formulates the cache-aided matrix multiplication  retrieval  problem.
Section~\ref{sec:main} summarizes the main results in this paper. 
Section~\ref{sec:achie} 
 provides the details of  the proposed coded cache-aided matrix multiplication  retrieval  schemes.
Section~\ref{sec:conclusion} concludes the paper.
Some proofs can be found in Appendix.

\subsection{Notation Convention}
\label{sub:notation}
Calligraphic symbols denote sets, 
bold symbols denote vectors and matrices,
and sans-serif symbols denote system parameters.
We use $|\cdot|$ to represent the cardinality of a set or the length of a vector;
$[a:b]:=\left\{ a,a+1,\ldots,b\right\}$ and $[n] := [1:n]$; 
$\oplus$ represents bit-wise XOR; 
$\mathbb{F}_{\qsf}$ represents a  finite field with order $\qsf$;         
$\mathbf{A}^{\text{\rm T}}$  and $\mathbf{A}^{-1}$ represent the transpose  and the inverse of matrix $\mathbf{A}$, respectively;
$\text{rank}(\mathbf{A})$ represents the rank of matrix $\mathbf{A}$; 
$\mathbf{I}_n$ represents the identity matrix  of  dimension $n \times n$;
$(\mathbf{A})_{m \times n}$ explicitly indicates that the matrix $\mathbf{A}$ is of dimension $m \times n$;
we let $\binom{x}{y}=0$ if $x<0$ or $y<0$ or $x<y$.
 In the rest of the paper entropies will be in base $\qsf$, where $\qsf$ will be introduced later.

\section{Preliminary Results on the Entropy of a Matrix Product}
\label{sec:pre}


In this section we describe a procedure to ``compress'' matrix products that may not be full rank so as to reduce the load on the shared-link.

Consider a matrix $\mathbf{A}\in \mathbb{F}_{\qsf}^{M \times m}$ on a finite field $\mathbb{F}_{\qsf}$ of rank $\rho$ with $M\geq m\geq \rho > 0$. 
We can choose $\rho$ linearly independent rows of $\mathbf{A}$ and call the resulting matrix $\mathbf{A}_1 \in \mathbb{F}_{\qsf}^{\rho \times m}$, that is, $\mathbf{A}_1 \mathbf{A}_1^{\text{\rm T}}\in \mathbb{F}_{\qsf}^{\rho \times \rho}$ is full rank.
We then can express the remaining $M-\rho$ rows of $\mathbf{A}$ as a linear combination of the rows of $\mathbf{A}_1$; 
let the matrix of the coefficients for the linear combinations be $\mathbf{A}_2 \in \mathbb{F}_{\qsf}^{(M-\rho) \times \rho}$.
Finally, the original matrix $\mathbf{A}$ can be written as $\mathbf{A} = \mathbf{A}_3 \begin{bmatrix} \mathbf{I}_\rho \\  \mathbf{A}_2 \\ \end{bmatrix} \mathbf{A}_1$, for some permutation matrix $\mathbf{A}_3  \in \{0,1\}^{M \times M} $ that only depends on the set of indices of the $\rho$ chosen rows out of $M$ rows. 
\begin{subequations} 
Thus we  can write
\begin{align}
H(\mathbf{A}) 
  &= H(\mathbf{A},  \mathbf{A}_1, \mathbf{A}_2, \mathbf{A}_3)  \label{eq:H-b1:all mats}
\\&= H(\mathbf{A}_1, \mathbf{A}_2, \mathbf{A}_2) + H(\mathbf{A}| \mathbf{A}_1, \mathbf{A}_2, \mathbf{A}_3) 
\\&= H(\mathbf{A}_1, \mathbf{A}_2, \mathbf{A}_3)
\\&\leq H(\mathbf{A}_1) + H(\mathbf{A}_2) +  H(\mathbf{A}_3)
\\&\leq \rho m  +(M-\rho)\rho +  \log_{\qsf} \left(\binom{M}{\rho} \right), 
\end{align}  
in other words, we need at most $(M+m)\rho - \rho^2$ symbols in $\mathbb{F}_{\qsf}$ to specify any $\mathbf{A}\in \mathbb{F}_{\qsf}^{M \times m}$ of rank $\rho$, up to a permutation matrix that contributes $\log_{\qsf} \left(\binom{M}{\rho} \right)$ to the entropy. 
\label{eq:H-b1}
\end{subequations}

Next, for any two matrices $\mathbf{C}\in \mathbb{F}_{\qsf}^{m \times n}$ and $\mathbf{B}\in \mathbb{F}_{\qsf}^{n \times p}$, the entropy bound in~\eqref{eq:H-b1}, together with 
\begin{align}
{\rm rank}[\mathbf{C} \mathbf{B}]
\leq \min( {\rm rank}[\mathbf{C}], {\rm rank}[\mathbf{B}] )
\leq \min(n,m,p),
\label{eq:H-b2}
\end{align}
implies that we need, up to some symbols needed to describe a permutation, at most $f(m,n,p) = f(p,n,m)$ symbols in $\mathbb{F}_{\qsf}$ to specify the matrix product $\mathbf{C}  \mathbf{B} \in \mathbb{F}_{\qsf}^{m \times p}$ 
where the function $f(m,n,p)$ is defined as
\begin{align}
f(m,n,p) &:= (m+p-\min(n,m,p))\min(n,m,p) 
=\begin{cases}
(m+p-n)n & \min(m,p) \geq n \\
 mp      & \min(m,p) \leq n \\
\end{cases}.
\label{eq:H-b3}
\end{align}

In the rest of the paper, we will use $P(\mathbf{C}, \mathbf{B})$ to denote the $f(m,n,p)$ symbols  in $\mathbb{F}_{\qsf}$ that specify the matrix product $\mathbf{C}  \mathbf{B}$, up to a permutation. 

Note that for each product $\mathbf{C} \mathbf{B}$ considered in formulated cache-aided matrix multiplication problem (which will be clarified later), we assume that $m= a_1 n$ and $p= a_2 n$, where $a_1,a_2$ are fixed positive numbers and $n \to \infty$. In this case of large matrices, we have $\frac{\log_{\qsf} \left( \binom{\max(m,p)}{\min(n,m,p)} \right)}{f(m,n,p)} \to 0$ for any field size $\qsf$. This is because 
\begin{align*}
\log_{\qsf} \left( \binom{\max(m,p)}{\min(n,m,p)} \right) \leq \log_{\qsf} \left(  \max(m,p) !   \right) \leq \underbrace{\frac{3}{2}\log_{\qsf}(e)+(\max(m,p)+\frac{1}{2})\log_{\qsf}(\frac{\max(m,p)}{e})}_{\text{by Stirling's approximation}}.
\end{align*}
Hence, we have 
\begin{align}
 \frac{\log_{\qsf} \left( \binom{\max(m,p)}{\min(n,m,p)} \right)}{f(m,n,p)} \leq \frac{\frac{3}{2}\log_{\qsf}(e)+(\max(a_1,a_2)n+\frac{1}{2})\log_{\qsf}(\frac{\max(a_1,a_2)n}{e})}{f(a_1 n,n, a_2 n)} \to 0,
\end{align}
which comes from that $a_1,a_2$ are fixed positive numbers and $n \to \infty$. So in this case of large matrices,  $P(\mathbf{C}, \mathbf{B})$  is enough to specify $\mathbf{C}  \mathbf{B}$.

In addition,~\cite[Lemma~2]{secure2019jia} proved that   for any two independent matrices $\mathbf{C}\in \mathbb{F}_{\qsf}^{m \times n}$ and $\mathbf{B}\in \mathbb{F}_{\qsf}^{n \times p}$ with uniformly i.i.d. entries on $\mathbb{F}_{\qsf}$ where $\qsf \to \infty$, we have 
\begin{align}
H( \mathbf{C}  \mathbf{B}) = f(m,n,p). \label{eq:entropy of product iid inf}
\end{align}
That is because, in this case we have with overwhelming probability that 
\begin{enumerate}
\item the matrices corresponding to $\mathbf{A}_1, \mathbf{A}_2$ in~\eqref{eq:H-b1:all mats} for the matrix $\mathbf{A} = \mathbf{C}\mathbf{B}\in \mathbb{F}_{\qsf}^{m \times p}$ have uniformly i.i.d. entries, which leads $H(\mathbf{A}_1, \mathbf{A}_2)=H(\mathbf{A}_1) + H(\mathbf{A}_2)=f(m,n,p)$;
\item  the first ${\rm rank}( \mathbf{C}  \mathbf{B})$ rows of $\mathbf{C}  \mathbf{B}$ are linearly independent, so that $\mathbf{A}_3$ is an identity matrix (i.e., $H(\mathbf{A}_3)=0$).
\end{enumerate} 
In this case, $P(\mathbf{C}, \mathbf{B})$  is also enough to specify $\mathbf{C}  \mathbf{B}$.

For later use, we express $f(m,n,p)=g\left( \frac{m}{n}, \frac{p}{n} \right) n^2$, 
where $g(\alpha,\beta)$ is a symmetric function in its arguments as is defined as
\begin{align}
g(\alpha,\beta):= 
\begin{cases}
 \alpha+\beta-1 &  \min(\alpha,\beta ) \geq 1 \\
 \alpha \beta   &  \min(\alpha,\beta ) \leq 1 \\
\end{cases}. 
\label{eq:def of g}
\end{align}
Note that $\frac{g(\alpha,\alpha)}{\alpha} \leq 2$ simply says that the entropy of the product of two  matrices is at most the sum of the entropies of the two matrices.

\section{System Model}
\label{sec:model}


The $(\Ksf,\Nsf,\asf)$ shared-link cache-aided matrix multiplication retrieval  problem  is defined as follows. 
A server has access to a library of $\Nsf$ matrices, denoted by $\mathbf{W}_1,\ldots,\mathbf{W}_{\Nsf}$, and each matrix is  of   dimension $\ssf \times \rsf$ over a finite field $\mathbb{F}_{\qsf}$, for some prime-power $\qsf$.
The column-row ratio of each  matrix is denoted by $\asf := \rsf / \ssf \in (0, \infty)$.
We further assume that each element of each matrix is uniformly i.i.d.  over $\mathbb{F}_{\qsf}$ and that $\qsf$ is sufficiently large so that   
the entropy of any matrix product $\mathbf{W}^{T}_i \mathbf{W}_j$ where $(i,j)\in [\Nsf]^2$ is 
\begin{align}
\Bsf := f(\rsf,\ssf,\rsf) =\ssf^2 g(\asf,\asf) \leq 2\rsf\ssf,  
\label{eq:definition of B}
\end{align}
 i.e., $\Bsf$ is the number of symbols in $\mathbb{F}_{\qsf}$ that suffices to specify any matrix product, as argued in Section~\ref{sec:pre}.\footnote{\label{foot:without assumption}
  Note that without the   assumption that $\qsf \to\infty$, each proposed achievable scheme  can still work  
   to let each user   retrieve its demanded matrix product. As showed in Section~\ref{sec:pre}, $\qsf \to \infty$ is needed for the converse of~\eqref{eq:entropy of product iid inf}, which characterizes the entropy of matrix product. In addition, this assumption is also needed for the proposed converse bounds on the  minimum worst-case load.  
  } 
The server is connected to $\Ksf$ users through an error-free shared link. 
The system operates as follows.

\paragraph{Placement Phase} 
During the cache placement phase, each user stores information about the $\Nsf$ matrices in its local cache without knowledge of future users' demands,
that is, there exist placement functions  $\phi_k, \ k\in[\Ksf]$, such that
\begin{align}
\phi_k &: \mathbb{F}_{\qsf}^{\Nsf \ssf \rsf} \to \mathbb{F}_{\qsf}^{\lfloor \Msf \ssf \rsf \rfloor}. 
\label{eq: placement functions def}
\end{align}
We denote the content in the cache of user~$k\in[\Ksf]$ by $Z_{k}=\phi_k(\mathbf{W}_1, \ldots, \mathbf{W}_\Nsf)$. 
The non-negative parameter $\Msf$ is the {\it cache size}, measured in multiple of the size of each matrix in the library. 

\paragraph{Delivery Phase} 
During the delivery phase, user~$k \in [\Ksf]$ sends its demand $\dv_k=(d_{k,1},d_{k,2})$ to the server, where $(d_{k,1}, d_{k,2}) \in [\Nsf]^2$ means that user~$k$ requests the matrix product $\mathbf{W}^{\text{\rm T}}_{d_{k,1}} \mathbf{W}_{d_{k,2}} \in \mathbb{F}_{\qsf}^{\rsf \times \rsf}$. 
Given the demand  $[\dv_1;\dv_2;\cdots;\dv_{\Ksf}]\in [\Nsf]^{2\times \Ksf}$,  the server broadcasts the message 
$ X = \psi([\dv_1;\dv_2;\cdots;\dv_{\Ksf}], \mathbf{W}_1,\ldots,\mathbf{W}_{\Nsf} ) $
to the users, where the encoding function $\psi$ is such that
\begin{align} 
 \psi &: [\Nsf]^{2 \Ksf} \times \mathbb{F}_{\qsf}^{ \Nsf\ssf\rsf } \to \mathbb{F}_{\qsf}^{\lfloor \Rsf\Bsf \rfloor},.
\label{eq: encoding function def}
\end{align}
The non-negative parameter $\Rsf$ is referred to as the {\it load} on the shared link, measured in multiple of the entropy of a matrix product $\Bsf$ defined in~\eqref{eq:definition of B}.

\paragraph{Correctness} 
Each user~$k\in [\Ksf]$ decodes its desired matrix product from $([\dv_1;\dv_2;\cdots;\dv_{\Ksf}],Z_k,X)$ through the decoding function $\xi_k$, defined as
\begin{align}
\xi_k &:  [\Nsf]^{2 \Ksf} \times \mathbb{F}_{\qsf}^{ \lfloor\Msf \ssf \rsf\rfloor} \times \mathbb{F}_{\qsf}^{ \lfloor\Rsf\Bsf\rfloor} \to \mathbb{F}_{\qsf}^{\Bsf},
\label{eq: decoding functions def}
\end{align}
such that  
\begin{align}
H\big(\mathbf{W}^{\text{\rm T}}_{d_{k,1}} \mathbf{W}_{d_{k,2}} | \xi_k ([\dv_1;\dv_2;\cdots;\dv_{\Ksf}],Z_k,X)\big) = 0, \ \forall k\in[\Ksf].
\label{eq: correctness def}
\end{align}

\paragraph{Objective}
In this paper, we assume that the computation power  of the  server and users  is unlimited.  Therefore, our focus is on the optimal tradeoff between communication cost and cache storage capacity.   
 More precisely, we aim to determine the {\it minimum worst-case load} among all possible demands, defined for $\Msf\in[0,\Nsf]$ as
\begin{align}
\Rsf^\star  :=  \liminf_{\min(\qsf,\ssf)\to\infty} 
\min_{\substack{
(\phi_k, k\in [\Ksf]) \text{ in~\eqref{eq: placement functions def}}, \\
\psi \text{ in~\eqref{eq: encoding function def}}, \\
(\xi_k, k\in [\Ksf]) \text{ in~\eqref{eq: decoding functions def}}
}} \  \max_{[\dv_1;\dv_2;\cdots;\dv_{\Ksf}]} 
\{\Rsf : \text{the condition in~\eqref{eq: correctness def} is satisfied}\}.
\end{align}

\paragraph{Uncoded Cache Placement}
If each user directly copies  some symbols of the $\Nsf$ matrices  into its cache, the cache placement is said to be {\it uncoded}. 
The minimum worst-case load under the constraint of uncoded cache placement is denoted by $\Rsf^{\star}_{\text{u}}$.

\paragraph{Isomorphic Demands} 
Since $W^{\text{\rm T}}_{i} W_j = \left( W^{\text{\rm T}}_{j} W_i \right)^{\text{\rm T}}$ for any $(i,j)\in [\Nsf]^2$, we say that the demands $W^{\text{\rm T}}_{i} W_j$  and  $ W^{\text{\rm T}}_{j} W_i$ are {\it isomorphic}. The number of non-isomorphic demands is $\binom{\Nsf}{2}+\Nsf=\frac{\Nsf(\Nsf+1)}{2}=\binom{\Nsf+1}{2}$.
In this paper,  without  loss of generality, we thus can  assume that $d_{k,1} \leq d_{k,2}$ for each $k\in [\Ksf]$.

\begin{rem}[Range of $\Msf$]
\label{rem:rangeM}
\rm 
Note that when $\Msf\geq  \min\left( \Nsf, \frac{\Nsf(\Nsf+1)}{2}  \frac{g(\asf,\asf) }{\asf }  \right) $, we have $\Rsf^\star=0$. Indeed, the server does not need to send anything if each user can either store
all possible matrices in the library (requiring $\Nsf\rsf \ssf$ symbols) or 
all possible non-isomorphic matrix products (requiring $ \frac{\Nsf(\Nsf+1)}{2}  \Bsf $ symbols). 
 Recall that $\frac{\Bsf}{\rsf \ssf} = \frac{g(\asf,\asf)}{\asf} = \min(\asf,2-1/\asf)$. 
Hence, only for $\Msf < \min\left( \Nsf, \frac{\Nsf(\Nsf+1)}{2}  \frac{g(\asf,\asf) }{\asf }  \right) $ the load may be non-zero,
in which case we have
$\Rsf^\star  \leq \min\left(\Ksf, \frac{\Nsf(\Nsf+1)}{2},  \Nsf \frac{\asf}{g(\asf,\asf)} \right),$
as the server can satisfy all requests by either sending all demanded non-isomorphic matrix products (requiring  $\min(\Ksf, \frac{\Nsf(\Nsf+1)}{2}) \Bsf$ symbols), or all matrices in the library (requiring $\Nsf\rsf \ssf$ symbols). 
\hfill $\square$ 
\end{rem}

\section{Main Results and Discussions}
\label{sec:main}

This Section is organized as follows.
We first summarize our main results in Section~\ref{sub:main}. 
We then provide two examples to illustrate the main ingredients of our novel achievable schemes in Section~\ref{sub:highlevel}. 
We provide some numerical evaluations in Section~\ref{sub:numerical}.
Finally,  we discuss the difference between the proposed cache-aided matrix multiplication retrieval schemes and the existing works on distributed matrix multiplication for straggler mitigation in Section~\ref{sub:discussions}.

\subsection{Main Results}
\label{sub:main}  
For the $(\Ksf,\Nsf,\asf)$  shared-link cache-aided matrix multiplication retrieval problem, a  simple  solution is to treat each non-isomorphic product as an independent file, and thus the considered problem becomes a coded caching problem for  single file  retrieval with $\Ksf$ users and $\frac{\Nsf(\Nsf+1)}{2}$ files, for which we can directly use  the MAN coded caching scheme for  single file  retrieval.
Such a scheme is agnostic of the structure of matrix multiplication, and thus we refer to it as {\it structure-agnostic scheme}.
The achieved load by the structure-agnostic scheme is given as follows. The proof can be found in Appendix~\ref{sub:agnostic}.
\begin{thm}[Structure-agnostic scheme]
\label{thm:load of agnostic}
For the $(\Ksf,\Nsf,\asf)$  shared-link cache-aided matrix multiplication retrieval  problem,   $\Rsf^{\star} \leq \Rsf_{\text{\rm sa}}$, where $\Rsf_{\text{\rm sa}}$ is the lower convex envelope of the following memory-load pairs 
\begin{align}
(\Msf, \Rsf_{\text{\rm sa}})=\left(\frac{\Nsf(\Nsf+1)}{2} \ \frac{g(\asf,\asf)}{\asf} \ \frac{t}{\Ksf}, \frac{\Ksf-t}{t+1} \right), \ t\in [0:\Ksf]. \label{eq:load of agnostic}
\end{align}
\end{thm}
Note that when $\Msf=\frac{\Nsf(\Nsf+1)}{2}  \frac{g(\asf,\asf) }{\asf }$, i.e., $t=\Ksf$, we have $\Rsf_{\text{\rm sa}}=0$---see also Remark~\ref{rem:rangeM}.

The structure-agnostic scheme does not perform well when $\Nsf$ is large, because the number of non-isomorphic matrix products increases quadratically with $\Nsf$.
We can improve on Theorem~\ref{thm:load of agnostic} by designing structure-aware caching schemes, which leverage the specific structure of matrix multiplication. 
In the structure-agnostic scheme, each user directly caches the elements in the matrix products; 
in the proposed structure-aware caching schemes, each user caches $\frac{\Msf}{\Nsf} \ssf\rsf$ symbols of each matrix in the library. 

We first introduce two baseline structure-aware schemes.
In the first baseline scheme, referred to as {\it uncoded caching baseline scheme}, each user caches $\frac{\Msf}{\Nsf} \rsf$ columns of each matrix in the library; thus 
each user  can reconstruct  $\left( \frac{\Msf}{\Nsf} \rsf \right)^2$ elements of each matrix product  from its cached content.
In the second baseline scheme, referred to as {\it mutli-request baseline scheme}, each user directly recovers the two library matrices instead of their product,  akin  
to a coded caching scheme with multiple file retrieval~\cite{multiJi2014}.
The achieved loads by the baseline structure-aware schemes are given as follows. 
The proof details can be found in Sections~\ref{sub:baseline 1} and~\ref{sub:baseline 2}, respectively.
\begin{thm}[Baseline structure-aware schemes]
\label{thm:base}
For the $(\Ksf,\Nsf,\asf)$  shared-link cache-aided matrix multiplication  retrieval  problem, $\Rsf^{\star} \leq \min(\Rsf_1,\Rsf_2) $ where 
$\Rsf_1$ is defines as 
\begin{align}
\Rsf_1:=  \Ksf\left( 1- \frac{\Msf^2}{\Nsf^2} \right) \frac{\asf^2}{g(\asf,\asf)}, 
\label{eq:first baseline}
\end{align}
and $\Rsf_2$ is the lower convex envelope of the following memory-load pairs 
\begin{align}
&(\Msf, \Rsf_2)= \left(\Nsf \frac{t}{\Ksf},  2\frac{\Ksf-t}{t+1} \ \frac{\asf }{g(\asf,\asf)} \right), \ \forall t \in [0:\Ksf] . \label{eq:second baseline} 
\end{align}
\end{thm}

The main limitation of the first baseline scheme in~\eqref{eq:first baseline} is the use of uncoded caching (i.e., there is no  multicasting gain). 
The main limitation of the second baseline scheme in~\eqref{eq:second baseline} is that it is not necessary to recover the two library matrices in order to recover their product.
In order to improve on the baseline structure-aware schemes, we next propose two 
schemes where we partition the matrices in the library into sub-matrices and then let a subset of the users cache  (a linear transformation  of)  each sub-matrix.
The achieved load of the row-partition scheme is given as follows.
The proof details can be found in Section~\ref{sub:first approach}.

\begin{thm}[Row-partition scheme]
\label{thm:first approach}
For the $(\Ksf,\Nsf,\asf)$  shared-link cache-aided matrix multiplication  retrieval  problem,  $\Rsf^{\star} \leq \Rsf_{\text{\rm row}}$, where   
\begin{subequations}
\begin{align}
&  \Rsf_{\text{\rm row}}:=\min_{\ell \in [\Ksf]} \left\lceil \frac{\Ksf}{\ell}\right\rceil  \frac{ g\left( \frac{\asf \binom{\ell}{t_{\ell}}}{\alpha_{\ell}},\frac{\asf \binom{\ell}{t_{\ell}}}{\alpha_{\ell}}\right) \frac{\alpha^2_{\ell}}{ \binom{\ell}{t_{\ell}}^2 } \binom{\ell}{t_{\ell}+1}+  g\left( \frac{\asf \binom{\ell}{t_{\ell}+1}}{1-\alpha_{\ell}}, \frac{\asf \binom{\ell}{t_{\ell}+1}}{1-\alpha_{\ell}}\right) \frac{(1-\alpha_{\ell})^2}{ \binom{\ell}{t_{\ell}+1}^2 }
  \binom{\ell}{t_{\ell}+2} }{  g(\asf,\asf)  }, 
  \label{eq:R_row}\\
&\alpha_{\ell}:= t_{\ell}+1 -\frac{\ell \Msf}{\Nsf}, \ \ell \in [\Ksf], 
\label{eq:def of alpha}\\
& t_{\ell}:= \left\lfloor \frac{ \ell \Msf}{\Nsf} \right\rfloor, \ \ell \in [\Ksf], 
\label{eq:def of t1}
\end{align}
%
with the convention that 
\begin{align}
 \Rsf_{\text{\rm row}} & = 
 \min_{\ell \in [\Ksf]} \left\lceil \frac{\Ksf}{\ell}\right\rceil  \frac{ g\left(  \asf \binom{\ell}{t_{\ell}} , \asf \binom{\ell}{t_{\ell}} \right)      \binom{\ell}{t_{\ell}+1} }{  g(\asf,\asf)  \ \binom{\ell}{t_{\ell}}^2 }  \text{ when $\alpha_{\ell}=1$, and }
\\
 \Rsf_{\text{\rm row}} & =    
 \min_{\ell \in [\Ksf]} \left\lceil \frac{\Ksf}{\ell}\right\rceil  \frac{  g\left( \asf \binom{\ell}{t_{\ell}+1} , \asf \binom{\ell}{t_{\ell}+1} \right)  
  \binom{\ell}{t_{\ell}+2} }{ g(\asf,\asf)  \ \binom{\ell}{t_{\ell}+1}^2 }   \text{ when $\alpha_{\ell}=0$}. 
\end{align}
\end{subequations} 
\end{thm}
In Remark~\ref{rem:all caching}  we shall argue  that the row-partition strategy in Theorem~\ref{thm:first approach} can be used with any known (for the shared-link caching problem for  single file  retrieval) caching scheme with uncoded cache placement.

The achieved load of the column-partition scheme is given as follows.
The proof details can be found in Section~\ref{sub:second approach}.
\begin{thm}[Column-partition scheme]
\label{thm:second approach}
For the $(\Ksf,\Nsf,\asf)$  shared-link cache-aided matrix multiplication  retrieval  problem, $\Rsf^{\star}\leq \Rsf_{\text{\rm col}}$, where  
\begin{subequations}
\begin{align}
&\Rsf_{\text{\rm col}}:= \begin{cases} y, & \text{ if }  \asf \leq   1; \\ \frac{y+2 (\asf-1) \left( \alpha_{\Ksf} \frac{\Ksf-t_{\Ksf}}{t_{\Ksf}+1} +(1-\alpha_{\Ksf}) \frac{\Ksf-t_{\Ksf}-1}{t_{\Ksf}+2}\right)}{2 \asf -1}   & \text{ if }  \asf >  1; \end{cases} \label{eq:R_col} \\
&   y:=  \sum_{i\in [0:t_{\Ksf}+1]} \binom{\Ksf}{i+1}  \left( 
\frac{\alpha_{\Ksf}^2 }{\binom{\Ksf}{t_{\Ksf}}^2} \binom{\Ksf-i}{t_{\Ksf}-i} \binom{\Ksf-t_{\Ksf}}{t_{\Ksf}-i }+ 
\frac{(1- \alpha_{\Ksf})^2 }{\binom{\Ksf}{t_{\Ksf}+1}^2}  \binom{\Ksf-i}{t_{\Ksf}+1-i} \binom{\Ksf-t_{\Ksf}-1}{t_{\Ksf}+1-i } 
\right. \nonumber\\ & \left. \qquad 
+ 2  \frac{\alpha_{\Ksf} (1- \alpha_{\Ksf})}{\binom{\Ksf}{t_{\Ksf}} \binom{\Ksf}{t_{\Ksf}+1}}   \binom{\Ksf-i}{t_{\Ksf}-i} \binom{\Ksf-t_{\Ksf}}{t_{\Ksf}+1-i } \right),\label{eq:def of y}
\end{align} 
\end{subequations} 
where  $t_{\Ksf}:= \left\lfloor \frac{ \Ksf \Msf}{\Nsf} \right\rfloor  \in[0:\Ksf] $   and  
   $\alpha_{\Ksf}= \left\lfloor \frac{ \Ksf \Msf}{\Nsf} \right\rfloor +1 -\frac{\Ksf\Msf}{\Nsf}  \in[0,1] $ were defined in~\eqref{eq:def of t1} and~\eqref{eq:def of alpha}, respectively. 
\end{thm}

In Remark~\ref{rem:proof cor formula 1} and Remark~\ref{rem:proof cor formula 2} we will show that the proposed row- and column-partition schemes outperform the two baseline schemes, respectively,  and therefore we have the following Corollary.
\begin{cor}
\label{cor:improve}
For the $(\Ksf,\Nsf,\asf)$  shared-link cache-aided matrix multiplication  retrieval  problem,  we have $\Rsf_{\text{\rm row}} \leq  \Rsf_2 $ and $\Rsf_{\text{\rm col}} \leq \Rsf_1$, for    all $\Msf \in [0,\Nsf]$.
\end{cor}

\begin{rem}[Structure-agnostic vs structure-aware  schemes]
\label{rem:agnostic can be better}
\rm 
We note that the proposed structure-aware schemes in this paper are not always better than the proposed structure-agnostic scheme. When $\asf$ is very small, the structure-agnostic scheme outperforms the other schemes, because in this case the dimension of each matrix product is much less than the input matrices and thus it is more efficient to directly cache the matrix products. For example, if 
   $\frac{\Nsf(\Nsf+1)}{2}  \frac{g(\asf,\asf) }{\asf }  < \Nsf  $ (i.e., $\asf<\frac{2}{\Nsf+1}$) and $\Msf=\frac{\Nsf(\Nsf+1)}{2}  \frac{g(\asf,\asf) }{\asf } $, 
   the achieved load of the structure-agnostic scheme is $0$ (see also Remark~\ref{rem:rangeM}), 
   while the achieved loads of the structure-aware schemes are  strictly  larger than $0$.
   
In general, see also Section~\ref{sub:numerical} for numerical evaluations, the row-partition scheme does not uniformly outperforms the column-partition scheme, or vice versa.
Thus, for the proposed schemes, we cannot infer any uniform superiority of a certain placement strategy. 
\hfill $\square$ 
\end{rem}

\begin{rem}[On redundant multicast messages]
\label{rem:YMA}
\rm
In this paper's proposed coded caching schemes, after generating the coded symbols desired by the users, we use the MAN delivery scheme to generate multicast messages to deliver those coded symbols. Yu, Maddah-Ali and Avestimehr  in~\cite{exactrateuncoded} showed that some MAN multicast messages  may be redundant when a file is requested by multiple users, and thus need not be transmitted. 
In our coded caching schemes, if there exist some products demanded by several users, we could use the approach in~\cite{exactrateuncoded} to remove the redundant multicast messages. We do not report here this type of enhancement for sake of conciseness.
\hfill $\square$ 
\end{rem}

\begin{rem}[Extensions]
\label{rem:extension}
\rm 
Similarly to~\cite[Remark 3]{arxivfunctionretrieval}, we can extend the proposed schemes to  Device-to-Device networks~\cite{d2dcaching}, where in the delivery phase  each user broadcasts coded  packets  based on its cached content   to all other users, and to the coded caching problem with private demands~\cite{wan2019privatecaching,wan2019fundamental}, where we aim to preserve the privacy of the demand of each user from other users. We do not report here this type of extensions for sake of conciseness.
\hfill $\square$ 
\end{rem}

So far we looked at achievable scheme. We now turn to converse bounds. We can directly use the cut-set bounds in~\cite{dvbt2fundamental,franarxiv} for the shared-link coded caching problem for single file retrieval into our problem, which leads to  the following theorem.
\begin{thm}[Cut-set converse bound]
\label{thm:cut-set converse}
For the $(\Ksf,\Nsf,\asf)$  shared-link cache-aided matrix multiplication  retrieval  problem,  we have
\begin{align}
  \Rsf^{\star}  \geq   \max_{b\in [\min( \Nsf^{\prime}, \Ksf ) ]}  \left( b  -  b^2 \frac{\Msf}{\Nsf^{\prime}} \frac{\asf}{g(\asf,\asf) } \right), \label{eq:cutset converse} 
\end{align}
where $\Nsf^{\prime}=\left\lfloor \frac{\Nsf}{2}  \right\rfloor $.
\end{thm}
\begin{IEEEproof} 
For each $i \in [\Nsf^{\prime}]$, define $W^{\prime}_i = \mathbf{W}^{T}_{2(i-1)+1} \mathbf{W}_{2 i}$. 
Consider a cut with $b \in [\min ( \Nsf^{\prime}, \Ksf )]$ users, and  let each user demand  one product $W^{\prime}_i$ where $i\in [\Nsf^{\prime}]$.
By using the   cut-set bound  in~\cite[Theorem 2]{dvbt2fundamental}, we have 
\begin{align}
 \left\lfloor \frac{\Nsf^{\prime}}{b}  \right\rfloor  \Rsf^{\star} \Bsf + b \Msf \ssf \rsf \geq  b  \left\lfloor \frac{\Nsf^{\prime}}{b}  \right\rfloor   \Bsf. \label{eq:MAN cut-set} 
\end{align}
Then, by using the strategy in~\cite[By-product 1]{franarxiv}, we can remove the `floor operator' in~\eqref{eq:MAN cut-set} and thus obtain~\eqref{eq:cutset converse}.
\end{IEEEproof}

When $\asf \geq 1$   and $\Nsf \geq 2\Ksf$, we propose a novel genie-aided converse bound  (proved in Appendix~\ref{sec:converse proof}) which smartly bounds the load by  the converse bound in~\cite{indexcodingcaching2020,exactrateuncoded} for the original MAN coded caching problem for single file retrieval. 
By using this novel   converse bound,  we have the following order optimality results.
\begin{thm}[Converse bound and order optimality result under uncoded cache placement]
\label{thm:order optimality}
For the $(\Ksf,\Nsf,\asf)$  shared-link cache-aided matrix multiplication  retrieval  problem where $\asf \geq 1$   and $\Nsf \geq 2\Ksf$,
 the worst-case load under the constraint of uncoded cache placement $\Rsf^{\star}_{\text{u}}$ is  
lower bounded by the lower convex envelop of 
\begin{align}
\left(\frac{\Nsf t}{\Ksf},  \frac{\Ksf-t}{t+1} \frac{ \ssf\rsf }{f(\rsf,\ssf, \rsf)}\right)=\left(\frac{\Nsf t}{\Ksf},  \frac{\Ksf-t}{t+1} \frac{ \asf }{2\asf -1 }\right), \ \forall t\in [0:\Ksf]. \label{eq:final converse equa}
\end{align}
In addition,  
 we have
 \begin{align}
 & \Rsf^{\star}_{\text{u}}  \geq  \frac{\Rsf_2}{2} \geq  \frac{\Rsf_{\text{\rm row}}}{2} \ \text{ when }  \asf \geq 1 \text{ and } \Nsf \geq 2\Ksf.
 \label{eq:uncoded order opt} 
 \end{align}
\end{thm} 
Note that the multiplicative gap between the converse bounds in Theorems~\ref{thm:cut-set converse} and~\ref{thm:order optimality} could be unbounded. For example, 
when $2\asf $ divides $\Ksf$ and    $\Msf = \frac{2\asf-1}{2\asf} \Nsf$, from Theorem ~\ref{thm:cut-set converse}  we have $\Rsf^{\star}\geq 0$ and from Theorem~\ref{thm:order optimality} we have $\Rsf^{\star}_{\text{u}}\geq  \frac{\asf \Ksf/(2\asf-1)}{\Ksf(2\asf-1)+2\asf} >0$. Hence, we cannot obtain the order optimality results in Theorem~\ref{thm:order optimality} from the cut-set converse bound in Theorem~\ref{thm:cut-set converse}.

\subsection{High-level Strategies for Theorems~\ref{thm:first approach} and~\ref{thm:second approach}}
\label{sub:highlevel}
In this section we provide two simple examples to highlight the key ideas in Theorems~\ref{thm:first approach} and~\ref{thm:second approach},  in which we partition each matrix in the library by columns and by rows, respectively.


\begin{example}[Case $\asf \leq 1$]\label{easy example 1: a>1}
In this example, there are $\Ksf=2$ users and $\Nsf=4$ matrices of dimension $\ssf \times \rsf=2 \times 2$ (i.e., $\asf =1$),  where each user can store up to $8$ symbols (i.e., $\Msf=2$). Denote the four matrices as
\begin{align}
\mathbf{A}  = \begin{bmatrix}
 a_1   & a_2 \\  
 a_3   & a_4  
\end{bmatrix},
\mathbf{B}  = \begin{bmatrix}
 b_1   & b_2 \\  
 b_3   & b_4  
\end{bmatrix},
\mathbf{C}  = \begin{bmatrix}
 c_1   & c_2 \\  
 c_3   & c_4  
\end{bmatrix},
\mathbf{D}  = \begin{bmatrix}
 d_1   & d_2 \\  
 d_3   & d_4  
\end{bmatrix}.
\end{align}
For the delivery phase, assume that user~$1$ demands $\mathbf{A}^{\text{\rm T}} \mathbf{B}$ and user~$2$ demands $\mathbf{C}^{\text{\rm T}} \mathbf{D}$, where 
\begin{subequations}
\begin{align}
 \mathbf{A}^{\text{\rm T}} \mathbf{B}  = \begin{bmatrix}
 a_1 b_1+  a_3 b_3 &  a_1 b_2+  a_3 b_4 \\  
 a_2 b_1+  a_4 b_3 &  a_2 b_2+  a_4 b_4 
\end{bmatrix}=: \begin{bmatrix}
 p_1 &  p_2 \\  
 p_3    &  p_4 
\end{bmatrix},
\label{eq:ex a>1 demand1}
\\
\mathbf{C}^{\text{\rm T}} \mathbf{D}  = \begin{bmatrix}
 c_1 d_1+  c_3 d_3 &  c_1 d_2+  c_3 d_4 \\  
 c_2 d_1+  c_4 d_3 &  c_2 d_2+  c_4 d_4 
\end{bmatrix}=: \begin{bmatrix}
 q_1 &  q_2 \\  
 q_3    &  q_4 
\end{bmatrix}.
\label{eq:ex a>1 demand2}
\end{align}
\end{subequations}

Next we compare the performances of our schemes. 
\begin{enumerate}

\item Structure-agnostic scheme: In Theorem~\ref{thm:load of agnostic}, we treat each matrix product as an independent file and use the MAN coded caching scheme for single file retrieval  (for the case of $\Ksf=2$ users, $\frac{\Nsf(\Nsf+1)}{2}=10$ files and cache size $\Msf=2$ files) to transmit $28/5=5.6$ symbols.

\item Column-partition scheme: 
here we let user~$1$ cache the first column of each matrix (e.g., $a_1$ and $a_3$  for the first file and similarly for the other files), and let user~$2$ cache the second column of each matrix  (e.g., $a_2$ and $a_4$  for the first file and similarly for the other files). 

Based on the cached content, $p_1$  in~\eqref{eq:ex a>1 demand1} can be reconstructed by user~$1$ and $q_4$ in~\eqref{eq:ex a>1 demand2} can be reconstructed by user~$2$. By~\eqref{eq:first baseline} of Theorem~\ref{thm:base}, the server transmits the remaining three symbols in the matrix product desired by each user, for a total of $6$ symbols.

Based on the cached content, we further note that user~$1$ requests $p_4$  in~\eqref{eq:ex a>1 demand1} that can be reconstructed by user~$2$, while user~$2$ requests $q_1$ in~\eqref{eq:ex a>1 demand2} that can be reconstructed by user~$1$. Thus the sever can transmit the coded symbol $p_4+q_1$. Thus, we totally transmit $5$ symbols (i.e., the server transmits $(p_2, p_3, p_4+q_1, q_2,q_3)$)  as in  Theorem~\ref{thm:second approach}.  

Note that in this scheme, each user directly recovers the desired ``sum of products'' symbols (e.g., $p_4 = a_2b_2 + a_4b_4$).

\item Row-partition scheme: 
here we use $\ell=\Ksf=2$, in which case the cache replication placement  in  Theorem~\ref{thm:first approach}  reduces to the MAN cache placement---the role of $\ell$  will be clarified further in Example~\ref{ex:row example} and Remark~\ref{rem:ell4vs2}.

By~\eqref{eq:second baseline} of Theorem~\ref{thm:base}, each user directly recovers the two matrices that make up its desired matrix product. In other words, during the delivery phase user~$1$ recovers $a_3,a_4,b_3,b_4$, which are cached by user~$2$, and user~$2$ recovers $c_1,c_2,d_1,d_2$ which are cached by user~$1$. Hence, the servers transmits $a_3+c_1, a_4+c_2, b_3+d_1, b_4+d_2$, totally $4$ symbols.

To improve on the above, we let user~$1$ cache the first row of each matrix (e.g., $a_1$ and $a_2$  for the first file and similarly for the other files), and let user~$2$ cache the second row of each matrix  (e.g., $a_3$ and $a_4$  for the first file and similarly for the other files). 
%
The server transmits
\begin{align}
 (a_3b_3 +c_1 d_1, \  a_3 b_4 +c_1 d_2, \  a_4 b_3 +c_2 d_1, \ a_4 b_4 +c_2 d_2), \label{eq:example in intro}
\end{align}
such that user~$1$ can recover $(a_3b_3,a_3 b_4,a_4 b_3 ,a_4 b_4) $ and user~$2$ can recover $(c_1 d_1,c_1 d_2,c_2 d_1,c_2 d_2)$. 

By leveraging the correlation of the elements in the products, we can further reduce the number of transmissions. Upon observing  that 
\begin{align}
 a_4 b_4= ( a_3 b_3)^{-1} (a_3 b_4)  (a_4 b_3), 
 \quad 
 c_2 b_2= (c_1 d_1)^{-1} (c_1 b_2) ( c_2 d_1),
\end{align}
thus we realize that do not need to transmit $a_4 b_4 +c_2 d_2$ in~\eqref{eq:example in intro}. Hence, we only need to transmit $3$ symbols as in  Theorem~\ref{thm:first approach}.

Note that in this scheme, each user recovers each individual term (e.g., $a_3b_3$ for user~$1$) in the ``sum of products'' symbols (e.g., $a_1b_1+a_3b_3$). 
\end{enumerate}
\end{example}

\begin{example}[Case $\asf > 1$]\label{easy example 1: a<1}  
In this case, the main steps of the row-partition scheme  remain the same  as in Example~\ref{easy example 1: a>1}. Hence, we only present the main ideas of the column-partition scheme.

Consider the case of  $\Ksf=2$ users, $\Nsf=4$ matrices of dimension $\ssf \times \rsf=2 \times 4$  (i.e., $\asf=2$), and where
each user can store up to $16$ symbols (i.e., $\Msf=2$).
Assume that the four matrices are $\mathbf{A}, \mathbf{B}, \mathbf{C}, \mathbf{D}$. 
\begin{subequations}
We express the matrix $\mathbf{A}$ as follows 
\begin{align}
\mathbf{A} &=
\begin{bmatrix} 
 a_1 & a_2  
&a_3 & a_4   \\  
 a_5 & a_6  
&a_7 & a_8  
\end{bmatrix}
= \mathbf{A}_1
\begin{bmatrix}
\mathbf{I}_2, \mathbf{A}_1^{-1}\mathbf{A}_2 
\end{bmatrix}, 
\\
&\mathbf{A}_1 := 
\begin{bmatrix} 
 a_1 & a_2  \\
 a_5 & a_6  \\
\end{bmatrix}, 
\mathbf{A}_2 := 
\begin{bmatrix} 
a_3 & a_4   \\  
a_7 & a_8   \\
\end{bmatrix},
\end{align}
where we assumed that block $\mathbf{A}_1$ is full rank (this is true with high probability when the filed size is large); same for the remaining matrices.
  Note that the general column-partition scheme described in Section~\ref{sub:second approach}  also works for the case where  $\mathbf{A}_1$ is not full rank; thus for arbitrary finite field, the proposed scheme also works.
\end{subequations}

\paragraph*{Placement phase}  
user~$1$ caches 
$\begin{bmatrix}
 a_1     \\  
 a_5      
\end{bmatrix}$
and $\mathbf{A}_1^{-1} 
\begin{bmatrix}
 a_3     \\  
 a_7      
\end{bmatrix},
$
and user~$2$ caches 
$\begin{bmatrix}
 a_2     \\  
 a_6      
\end{bmatrix}$
and 
$\mathbf{A}_1^{-1} 
\begin{bmatrix}
 a_4     \\  
 a_8      
\end{bmatrix};
$
similarly for the other matrices.
Hence, each user caches $4$ symbols from each matrix; thus each user caches $16$ symbols in total.

\paragraph*{Delivery phase} 
Assume that the users~$1$ and~$2$ demand 
\begin{align}
\mathbf{A}^{\text{T}} \mathbf{B} = \left[\begin{array}{c:c}
 \mathbf{A}_1^{\text{T}}  \mathbf{B}_1  &  \mathbf{A}_1^{\text{T}}  \mathbf{B}_2 \\  \hdashline
 \mathbf{A}_2^{\text{T}}  \mathbf{B}_1  &  \mathbf{A}_2^{\text{T}}  \mathbf{B}_2
\end{array} \right]
, \quad
\mathbf{C}^{\text{T}} \mathbf{D} = \left[\begin{array}{c:c}
 \mathbf{C}_1^{\text{T}}  \mathbf{D}_1  &  \mathbf{C}_1^{\text{T}}  \mathbf{D}_2 \\  \hdashline
 \mathbf{C}_2^{\text{T}}  \mathbf{D}_1  &  \mathbf{C}_2^{\text{T}}  \mathbf{D}_2
\end{array} \right],
\end{align}
respectively, where each matrix product contains $4$ blocks.
The delivery phase of the column-partition scheme contains three steps:
\begin{itemize}

\item In the first step, we let user~$1$ recover $ \mathbf{A}_1^{\text{T}}  \mathbf{B}_1$ and let user~$2$ recover  $ \mathbf{C}_1^{\text{T}}  \mathbf{D}_1$. The delivery is exactly the same as the column-partition scheme in the previous example for $\asf \leq 1$. Thus, we need to transmit $5$ symbols.

\item In the second step, we let user~$1$ and user~$2$ recover 
\begin{align}
\mathbf{A}_1^{\text{T}}  \mathbf{B}_2=  \left[\begin{array}{c:c} \mathbf{A}_1^{\text{T}}\begin{bmatrix}
 b_3     \\  
 b_7      
\end{bmatrix} &   \mathbf{A}_1^{\text{T}}\begin{bmatrix}
 b_4     \\  
 b_8     
\end{bmatrix}\end{array} \right],
\quad
\mathbf{C}_1^{\text{T}}  \mathbf{D}_2 = 
 \left[\begin{array}{c:c}
\mathbf{C}_1^{\text{T}}\begin{bmatrix}
 d_3     \\  
 d_7      
\end{bmatrix} &  \mathbf{C}_1^{\text{T}}\begin{bmatrix}
 d_4     \\  
 d_8     
\end{bmatrix}\end{array} \right],
\end{align} 
respectively. 
Since user~$1$ has recovered $\mathbf{A}_1^{\text{T}} \mathbf{B}_1$ in the first step and cached $\mathbf{B}_1^{-1} 
\begin{bmatrix}
 b_3     \\  
 b_7      
\end{bmatrix}
$, it can compute  $\mathbf{A}_1^{\text{T}} \mathbf{B}_1 \mathbf{B}_1^{-1} 
\begin{bmatrix}
 b_3     \\  
 b_7      
\end{bmatrix} = \mathbf{A}_1^{\text{T}} \begin{bmatrix}
 b_3     \\  
 b_7      
\end{bmatrix}$.  Similarly, user~$2$ can recover $ \mathbf{C}_1^{\text{T}}\begin{bmatrix}
 d_3     \\  
 d_7      
\end{bmatrix}$. In addition, 
$ 
\mathbf{A}_1^{\text{T}} \begin{bmatrix}
 b_4     \\  
 b_8     
\end{bmatrix} = \mathbf{A}_1^{\text{T}} \mathbf{B}_1 \mathbf{B}_1^{-1}  \begin{bmatrix}
 b_4     \\  
 b_8     
\end{bmatrix},
$ 
where  $F^{\prime}_{1,\{2\}}:= \mathbf{B}_1^{-1}  \begin{bmatrix}
 b_4     \\  
 b_8     
\end{bmatrix} $ is    cached by user~$2$.  Similarly, $F^{\prime}_{2,\{1\}}:= \mathbf{D}_1^{-1}  \begin{bmatrix}
 d_3     \\  
 d_7     
\end{bmatrix} $ is requested by user~$2$ and  is cached by user~$1$. We let the server transmit   $F^{\prime}_{1,\{2\}}+ F^{\prime}_{2,\{1\}}$ for a total of $2$ symbols.

\item In the third step, we let user~$1$ recover 
$$
 \mathbf{A}_2^{\text{T}}  \mathbf{B}_1  =\left(\mathbf{A}^{-1}_1 \mathbf{A}_2  \right)^{\text{T}}  \mathbf{A}_1^{\text{T}} \mathbf{B}_1 \ \text{ and } \     \mathbf{A}_2^{\text{T}}  \mathbf{B}_2 = \left(\mathbf{A}^{-1}_1 \mathbf{A}_2  \right)^{\text{T}}  \mathbf{A}_1^{\text{T}} \mathbf{B}_2,
$$
and  let user~$2$ recover 
$$
 \mathbf{C}_2^{\text{T}}  \mathbf{D}_1  =\left(\mathbf{C}^{-1}_1 \mathbf{C}_2  \right)^{\text{T}}  \mathbf{C}_1^{\text{T}} \mathbf{D}_1 \ \text{ and } \     \mathbf{C}_2^{\text{T}}  \mathbf{D}_2 = \left(\mathbf{C}^{-1}_1 \mathbf{C}_2  \right)^{\text{T}}  \mathbf{C}_1^{\text{T}} \mathbf{D}_2.
$$
Note that  $\mathbf{A}_1^{\text{T}} \mathbf{B}_1 $ and $\mathbf{A}_1^{\text{T}} \mathbf{B}_2$ have been recovered by user~$1$; in addition, we have 
 $$\mathbf{A}^{-1}_1 \mathbf{A}_2 =   
 \left[\begin{array}{c:c}
\mathbf{A}^{-1}_1 \begin{bmatrix}
 a_3     \\  
 a_7      
\end{bmatrix} &  \mathbf{A}^{-1}_1 \begin{bmatrix}
 a_4     \\  
 a_8     
\end{bmatrix}\end{array} \right]
 ,
 $$ 
 where  $\mathbf{A}^{-1}_1 \begin{bmatrix}
 a_3     \\  
 a_7      
\end{bmatrix}$ is cached by user~$1$ and $F^{\prime\prime}_{1,\{2\}}:= \mathbf{A}^{-1}_1 \begin{bmatrix}
 a_4     \\  
 a_8     
\end{bmatrix}$ is cached by user~$2$.
 Similarly, user~$2$ only needs to recover $F^{\prime\prime}_{2,\{1\}} = \mathbf{C}^{-1}_1 \begin{bmatrix}
 c_3     \\  
 c_7     
\end{bmatrix}.$  We let the server transmit   $F^{\prime\prime}_{1,\{2\}}+ F^{\prime\prime}_{2,\{1\}}$ for a total of $2$ symbols.
\end{itemize}
Thus, the server transmits $5+2+2=9$ symbols in total. 
Had we directly used the column-partition scheme for the case $\asf \leq 1$, the server would have sent $5$ symbols for each block, for a total of $20$ symbols.
\end{example}

To conclude, the high-level ideas for the row-partition and the column-partition schemes,  as well as, their main advantages and limitations,  are as follows:
\begin{enumerate}

\item {\it Row-partition scheme.}
The first approach partitions each matrix by rows and use the cache replication strategy in~\cite{finiteanalysis}. It will be explained in Remark~\ref{rem:ell4vs2} that, the  cache replication strategy in the shared-link caching problem  for single file retrieval  aims to reduce the sub-packetization level compared to the MAN scheme. In our context, the proposed   cache replication strategy with row-partition reduces both the load and the sub-packetization level simultaneously.

The matrix product desired by each user can be expressed by a sum of products of sub-matrices. By further encoding each term in the sum into a coded packet with length equal to its entropy, we then use the MAN delivery scheme  to transmit the coded packets.

\item {\it Column-partition scheme.}
The second approach partitions each matrix by columns. We separately consider the case $\asf \leq 1$ and the case $\asf >1$. When $\asf \leq 1$, we  use the MAN cache placement  strategy in~\cite{dvbt2fundamental} and propose a multi-round delivery scheme to transmit the coded packets.   When $\asf >1$, 
each demanded matrix product is not full rank; thus
the entropy of each product is $(2\asf-1)\ssf^2$ which is strictly  less than the number of its elements $\asf^2 \ssf^2$, i.e.,  there exist some redundant elements in each product.   Hence, we partition each matrix in the library into two blocks, where the   cache placement of the first block is as in the MAN scheme and we propose to use a  coded cache placement for  the second block. In the delivery phase, each product is also partitioned into blocks and the correlation among blocks is taken into consideration during the encoding procedure.

\item {\it On types of placement.} 
We also remark that the structure-agnostic uses an inter-file coded placement, where coding occurs across the symbols of all files (i.e., matrices). 
The row-partition scheme and the column-partition scheme for $\asf \leq 1$ use uncoded cache placement.
Finally, the column-partition scheme for $\asf >1$ uses an intra-file coded placement, where coding only occurs within the symbols of the same file. 

\item {\it Advantages and limitations.} 
The main advantages and limitations of the proposed schemes are (see also Remarks~\ref{rem:ell4vs2} and~\ref{rem:row vs column}):
\begin{itemize}
\item {\it Row-partition scheme.}
Its main advantage is that multicast opportunities are fully leveraged. 
In other words, if we need to transmit a requested symbol to a user and this symbol is cached by $t$ other users, it is encoded in a multicast message with $t+1$ symbols,  where each symbol is cached by $t$ users and demanded by one user. 
However, each element in a desired matrix product is the sum of some products of the elements in the library matrices. 
The main limitation of the row-partition scheme is that each user recovers each individual product in the sum. 
\item {\it Column-partition scheme.} 
Its main advantage is to let each user directly recover each element in the desired matrix product. 
Its main limitation is that multicast opportunities are not fully leveraged.
\end{itemize}
\item {\it Open problems.}
In Theorem~\ref{thm:order optimality}, we show that the proposed schemes are order optimal under uncoded cache placement for the case where $\asf \geq 1$ and $\Nsf \geq 2\Ksf$. For the remaining cases, in particular for the case  $\asf <1$, 
it is part of our on-going works to improve the proposed row-partition and column-partition
schemes. This may be attained by using inter-file coded placements and by a new partition
approach that has both the advantages of the row-partition and of the column-partition schemes,
and overcome their limitations. The derivation of a non-trivial converse bound for this case is also part of on-going works. 
 \end{enumerate}

  \begin{figure}
    \centering
    \begin{subfigure}[t]{0.5\textwidth}
        \centering
        \includegraphics[scale=0.5]{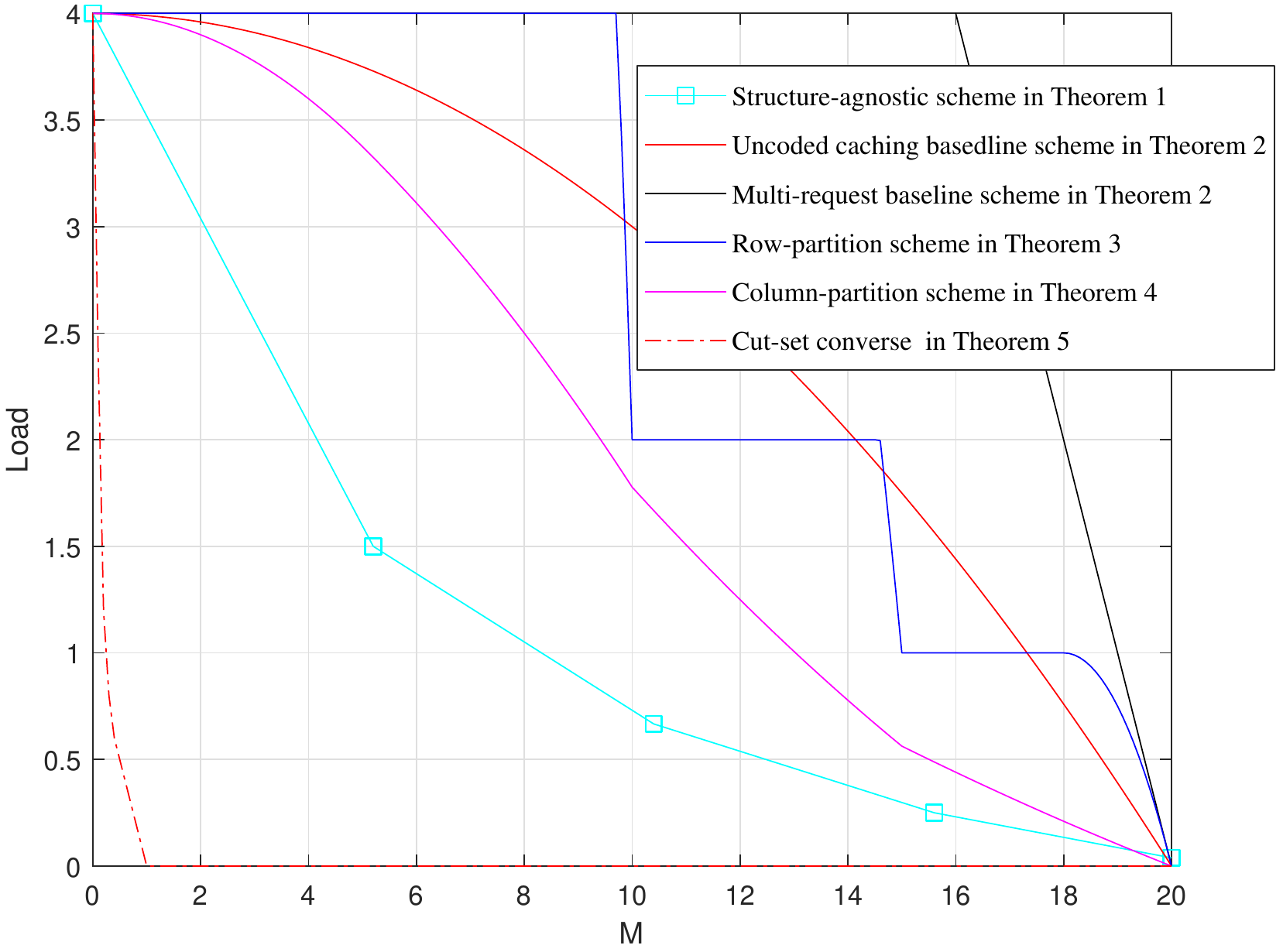}
        \caption{\small $\asf=1/10$.}
        \label{fig:numerical 1a}
    \end{subfigure}%
    ~ 
    \begin{subfigure}[t]{0.5\textwidth}
        \centering
        \includegraphics[scale=0.5]{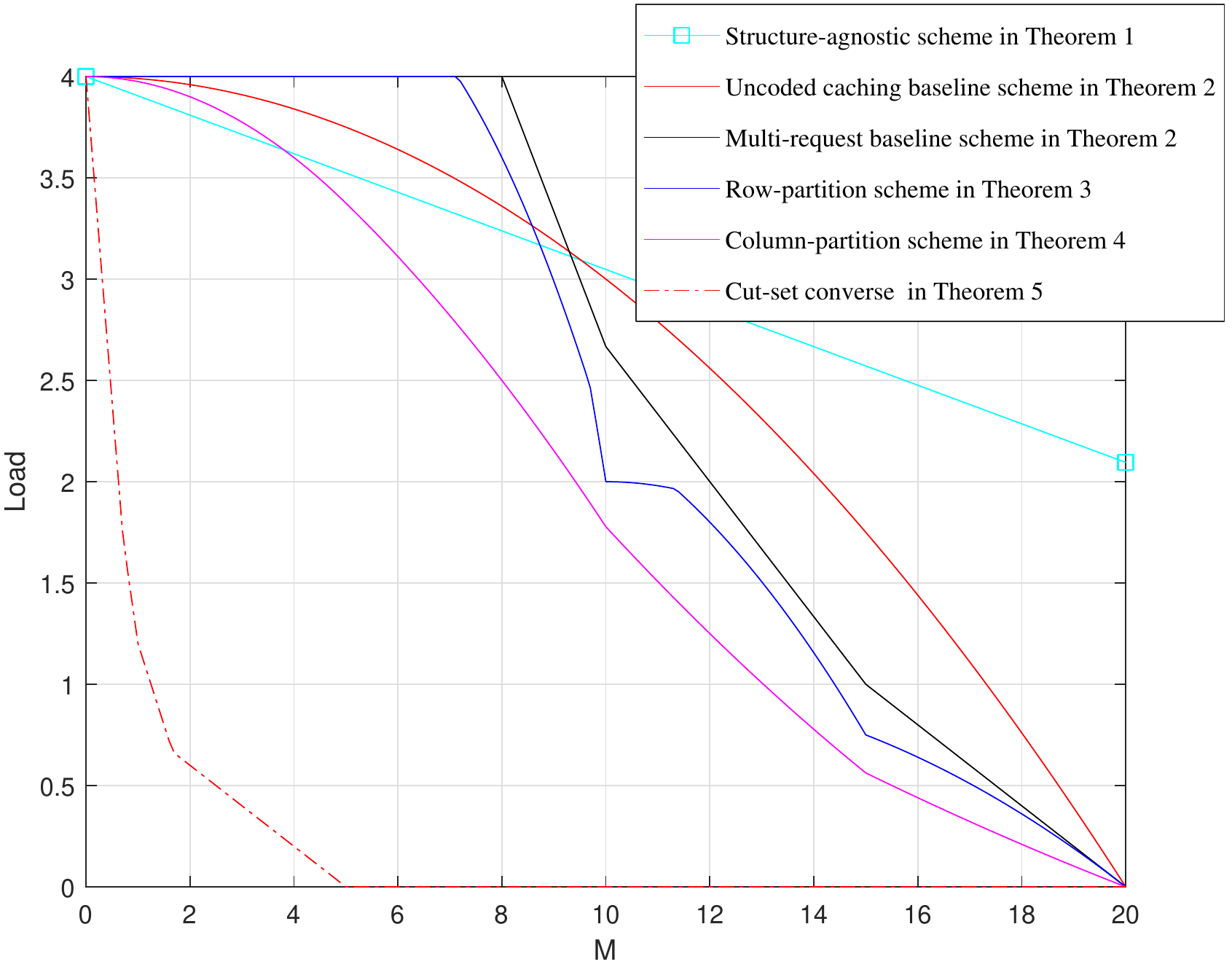}
        \caption{\small $\asf=1/2$.}
        \label{fig:numerical 1b}
    \end{subfigure}
   \\
       \centering
    \begin{subfigure}[t]{0.5\textwidth}
        \centering
        \includegraphics[scale=0.5]{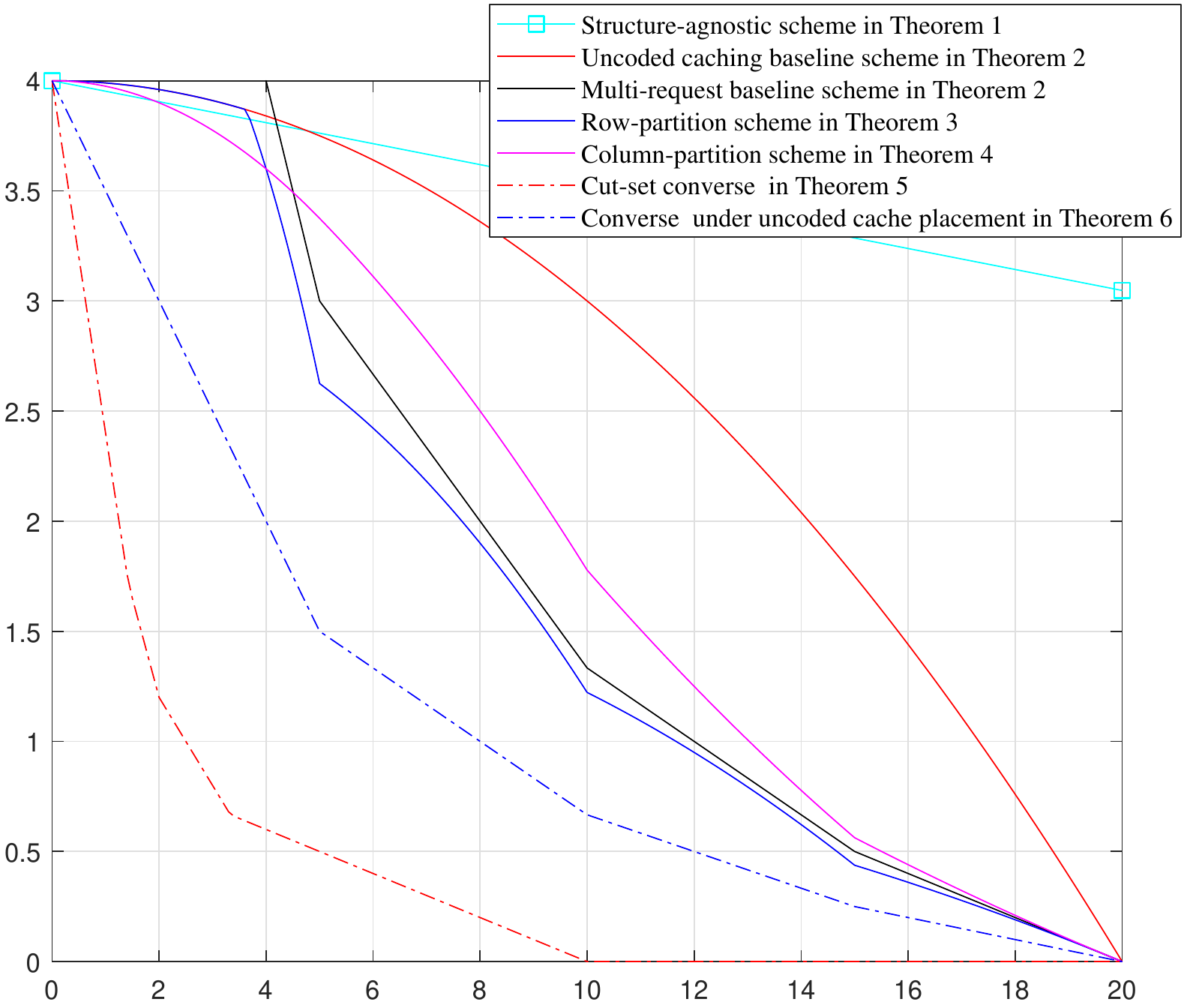}
        \caption{\small $\asf=1 $.}
        \label{fig:numerical 1c}
    \end{subfigure}%
    ~ 
    \begin{subfigure}[t]{0.5\textwidth}
        \centering
        \includegraphics[scale=0.5]{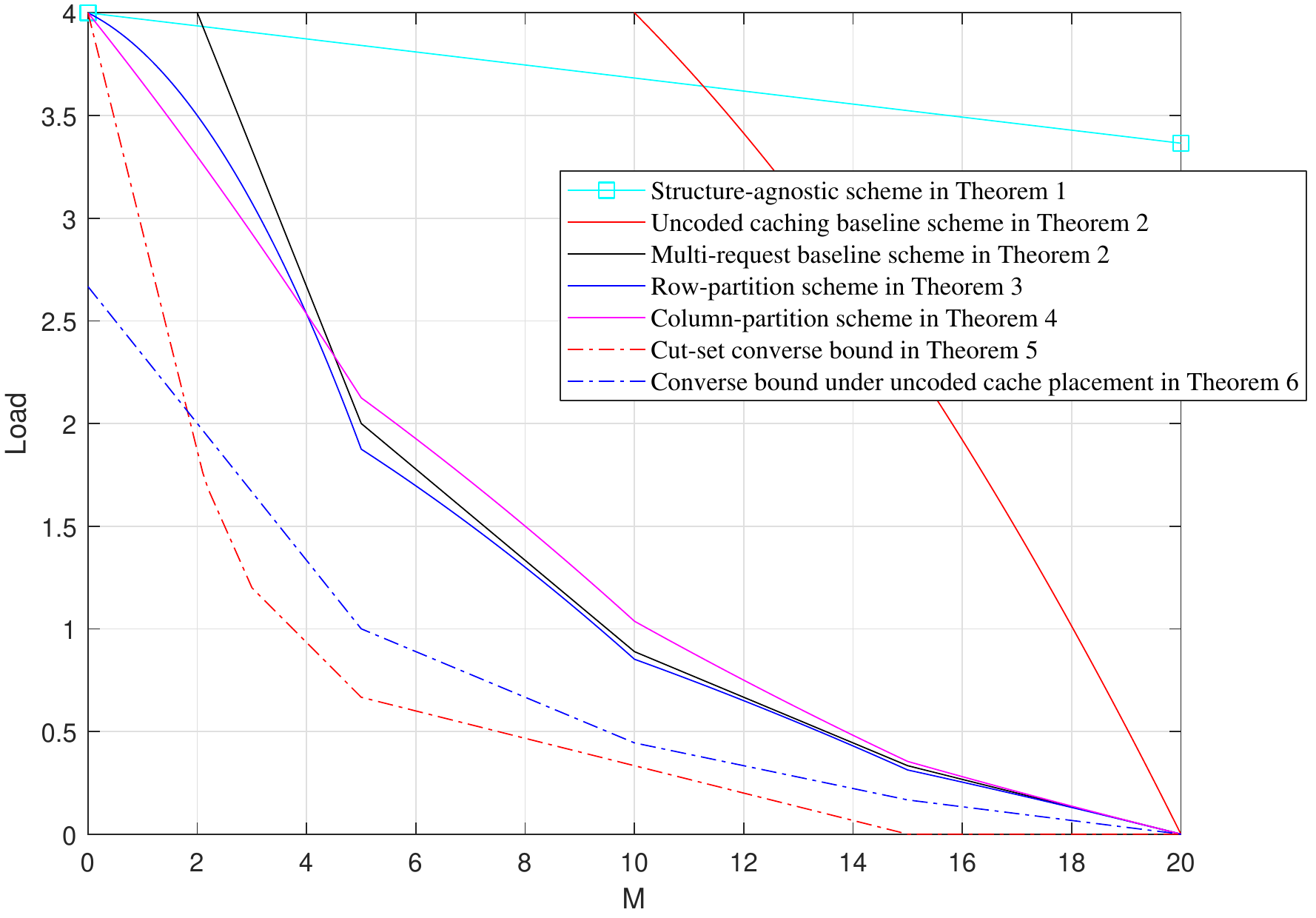}
        \caption{\small $\asf=2$.}
        \label{fig:numerical 1d}
    \end{subfigure}
    \\
        \centering
    \begin{subfigure}[t]{0.5\textwidth}
        \centering
        \includegraphics[scale=0.5]{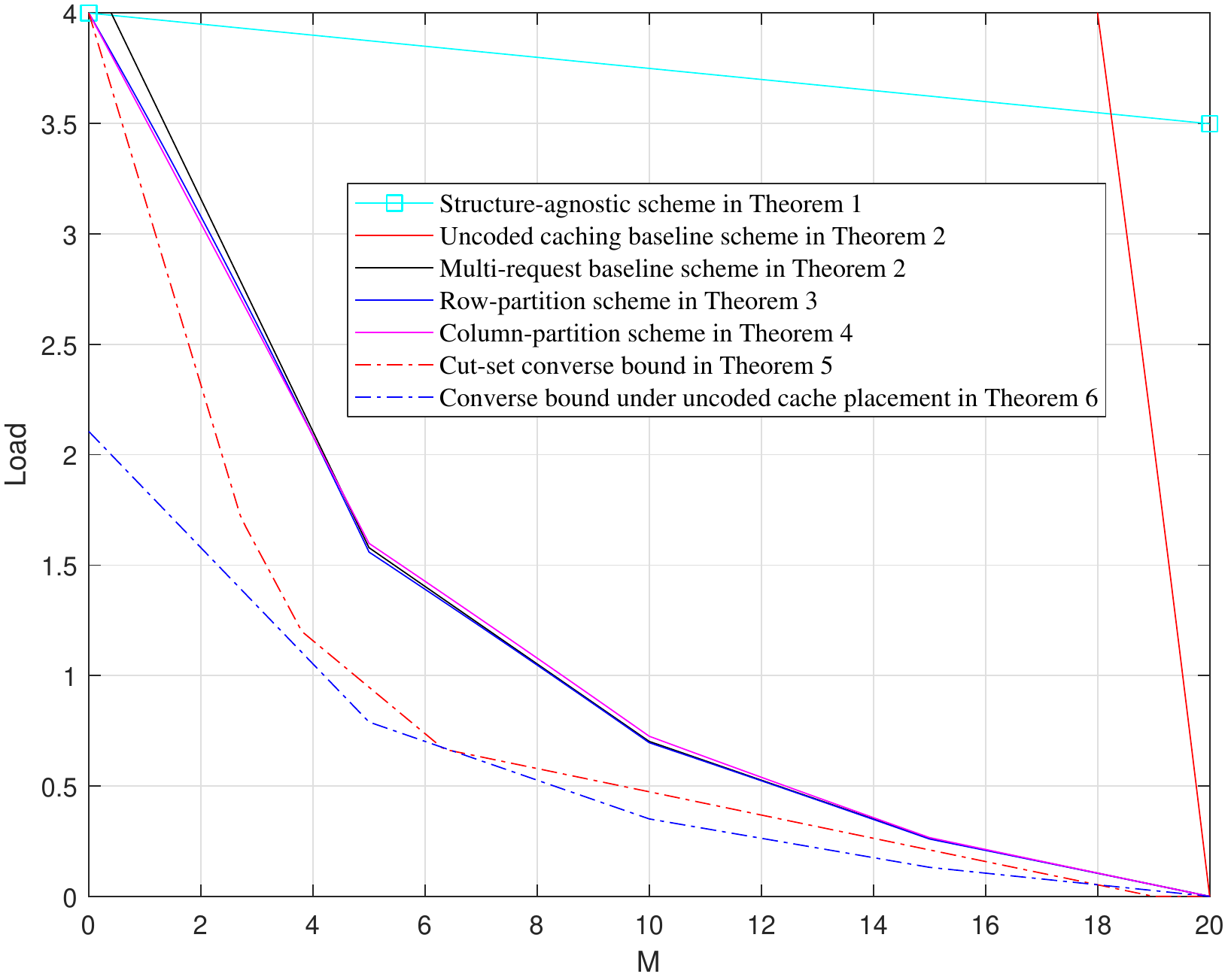}
        \caption{\small $\asf=10$.}
        \label{fig:numerical 1e}
    \end{subfigure}%
    \caption{\small Performance of various schemes for the shared-link cache-aided matrix multiplication retrieval problem with $\Ksf=4$ users and $\Nsf=20$ files for various values of the ratio $\asf$.}
    \label{fig:numerical 1}
\end{figure}

\subsection{Numerical Evaluations}
\label{sub:numerical}
We now provide some numerical evaluations for the proposed schemes and converse bounds.
In Fig.~\ref{fig:numerical 1}, we consider the case of $\Ksf=4$ users, $\Nsf=20$ files, and ratio $\asf \in \left\{ \frac{1}{10}, \frac{1}{2}, 1, 2 ,10 \right\}$. 
We observe the following from Fig.~\ref{fig:numerical 1}.
\begin{enumerate}

\item The row-partition scheme is always better than the multi-request baseline scheme, and the column-partition scheme is always better than the uncoded caching baseline scheme,  as Corollary~\ref{cor:improve} shows. 

\item When $\asf$ is small, the performance of the multi-request baseline scheme is much worse than the proposed row-partition and column-partition schemes. This is because in the multi-request baseline scheme each user recovers the two library matrices of its desired matrix product, which has $2\rsf\ssf$ symbols while the desired matrix product only has $\rsf^2$ symbols, which is  much lower than $2\rsf\ssf$ when $\asf$ is small.  

\item When $\asf$ is large, the performance of the uncoded caching baseline scheme is much worse than the proposed row-partition and column-partition schemes. This is because in the uncoded caching baseline scheme each user recovers all the $\rsf^2$ symbols in the desired matrix product. However, when $\asf$ is large, $\rsf^2$ is much larger than $f(\rsf,\ssf,\rsf)=2\ssf\rsf-\ssf^2$, which is the entropy of the matrix product.  

\item The structure-agnostic scheme performs well when $\asf$ is very small, since in this regime the entropy of each matrix product is much less than the entropy of each library matrix, and thus it is better to let the users directly cache the products. 

\item The load v.s. cache size curves
 may  not be convex. This is because in our setting we cannot memory-share between any two memory-load tradeoff points.  For example, if we partition each matrix in the library into two parts and use  a different  cache placement strategy on each part, in the product of any two matrices there may exist some elements computed  from both parts.   In this case the computation of the matrix multiplication cannot be divided into two separate parts, each of which is based on one cache placement strategy. 

\end{enumerate}

\subsection{Comparison to existing distributed matrix multiplication  computation  schemes for straggler mitigation} 
\label{sub:discussions}

The distributed matrix multiplication problem has received much attention in the recent years. 
The problem is as follows. 
There are two uniformly i.i.d. matrices $\mathbf{A}$  of  dimension $s \times r$ and $\mathbf{B}$  of  dimension $s \times t$, where $s \geq \min ( r,t)$. 
The matrix product $\mathbf{A}^{\text{\rm T}} \mathbf{B}$  must  be computed distributidely by a group of workers. 
There are mainly three strategies proposed in the literature, which partition each matrix into sub-matrices by rows~\cite{polyoptYu2017}, or by columns~\cite{dutta2020optimalRT}, or by blocks~\cite{yu2020stragglermitigation}.   
Each worker stores a linear combination of the sub-matrices in each matrix, and then computes the product of the two stored matrices, which is then sent to the master. From the transmissions of any $T$ workers, the master must be able to  correctly  recover the matrix product. The objective is to characterize the minimum $T$,  referred to as recovery threshold.

There are two main differences between our problem and the distributed matrix multiplication problem:
\begin{enumerate}

\item 
In our problem, there are multiple users receiving packets from the server, each of which caches some contents  from  the library and desires a product of two matrices. Hence, our problem is a {\it broadcast problem with side information}. By careful design, we aim to maximize the local caching gain (i.e., if some elements in the desired matrix product  have already been cached, we need not transmit them in the delivery phase) and  the coded caching multicasting gain. In contrast, in the distributed matrix multiplication  computation  problem, only the  master  wants to retrieve a product (no multicasting gain) and this master should recover the product only from the receiving packets (no local caching gain).

\item In the distributed matrix multiplication  computation  problem, it is assumed that  $s \geq \min ( r,t)$ (i.e., $\mathbf{A}^{\text{\rm T}} \mathbf{B}$ is full rank). Hence, each element in the  product    $\mathbf{A}^{\text{\rm T}} \mathbf{B}$ is also uniformly i.i.d. over $\mathbb{F}_{\qsf}$. The existing schemes let the master recover each element in the product individually (without leveraging the correlation among the elements in the product). Instead,   
our proposed schemes for this case (i.e., $\asf \leq 1$) still leverage the correlation among the elements in each product (see Example~\ref{easy example 1: a>1}). 
This is possible because each user cached some elements  of  each library matrix,  and with this side information its desired product could be further compressed.


\end{enumerate}


\section{Novel Structure-aware Achievable Schemes}
\label{sec:achie}

\subsection{Uncoded Caching Baseline Scheme: Proof of~\eqref{eq:first baseline}}
\label{sub:baseline 1}

\paragraph*{{ Placement phase}}
Each user caches the first $\frac{\Msf}{\Nsf} \rsf$ columns of each of the $\Nsf$ matrices in the library.  

\paragraph*{{ Delivery phase}}
User~$k \in [\Ksf]$ demands  $\mathbf{W}^{\text{\rm T}}_{d_{k,1}} \mathbf{W}_{d_{k,2}}$. Note that the first $\frac{\Msf}{\Nsf} \rsf$ rows of $\mathbf{W}^{\text{\rm T}}_{d_{k,1}} $ and the first $\frac{\Msf}{\Nsf} \rsf$ columns of $\mathbf{W}_{d_{k,1}}$ are cached by user~$k \in [\Ksf]$. Hence, user~$k \in [\Ksf]$ can directly recover  $  \frac{\Msf^2 \rsf^2}{ \Nsf^2 }$ elements of  $\mathbf{W}^{\text{\rm T}}_{d_{k,1}} \mathbf{W}_{d_{k,2}}$. Then we let the server directly transmit  the remaining $ \left( 1- \frac{\Msf^2}{\Nsf^2} \right)\rsf^2$ elements of $\mathbf{W}^{\text{\rm T}}_{d_{k,1}} \mathbf{W}_{d_{k,2}}$.
Hence, the total load is  
$$
\Ksf\left( 1- \frac{\Msf^2}{\Nsf^2} \right) \frac{\rsf^2}{f(\rsf,\ssf,\rsf)}=\Ksf\left( 1- \frac{\Msf^2}{\Nsf^2} \right) \frac{\asf^2}{g(\asf,\asf)},
$$
 which coincides  with~\eqref{eq:first baseline}.

\subsection{Multi-request Baseline Scheme: Proof of~\eqref{eq:second baseline}}
\label{sub:baseline 2}
We treat each matrix in the library as a file with $\ssf \rsf$ symbols, and use the coded caching scheme  for  multiple file retrieval  in~\cite{multiJi2014}.
We focus on each cache size $\Msf =\frac{\Nsf t}{\Ksf}$, where $t \in [0:\Ksf]$. 
 \paragraph*{{ Placement phase}}
We divide the  $\ssf \rsf$ symbols of each matrix $\mathbf{W}_{i}$ into $\binom{\Ksf}{t}$ non-overlapping and equal-length subfiles, $\mathbf{W}_{i}=\{W_{i,\Tc}: \Tc \subseteq [\Ksf], |\Tc|=t \}$. Each subfile $W_{i,\Tc}$ contains $\frac{\ssf \rsf}{ \binom{\Ksf}{t}}$ symbols 
and is cached exclusively by the users in $\Tc$.

 \paragraph*{{ Delivery phase}}
 User~$k \in [\Ksf]$ demands  $\mathbf{W}^{\text{\rm T}}_{d_{k,1}} \mathbf{W}_{d_{k,2}}$.
 We let user~$k \in [\Ksf]$ recover $\mathbf{W}_{d_{k,1}}$ and $\mathbf{W}_{d_{k,2}}$.
 For each set  $\Sc \subseteq [\Ksf]$ where $|\Sc|=t+1$,
 we let the server  broadcast the pair of multicast messages 
\begin{align}
&\sum_{k \in \Sc} W_{d_{k,1},\Sc \setminus \{k\}},
\quad
\sum_{k \in \Sc} F_{d_{k,2},\Sc \setminus \{k\}}. 
\label{eq:multi request multicast}
\end{align}
In $\sum_{k \in \Sc} W_{d_{k,1},\Sc \setminus \{k\}} $, user~$k$ stores all subfiles except $W_{d_{k,1},\Sc \setminus \{k\}}$ and thus it can recover this subfile. Similarly, user~$k$ can recover $W_{d_{k,2},\Sc \setminus \{k\}}$ from~\eqref{eq:multi request multicast}.
 
After considering all sets of users with cardinality $t+1$, each user can recover the two library matrices of its desired matrix product. The total load is  
$$
2 \binom{\Ksf}{t+1}  \frac{ \ssf \rsf }{\binom{\Ksf}{t}} \frac{1}{ f(\rsf,\ssf,\rsf)} =
 \frac{2(\Ksf-t) \asf }{(t+1)g(\asf,\asf)}, 
$$
which coincides  with~\eqref{eq:second baseline}.

\subsection{Row-partition Scheme: Proof of Theorem~\ref{thm:first approach}}
\label{sub:first approach}
We will start with a more detailed  example than the one in Section~\ref{sub:highlevel} to introduce the row-partition scheme in Theorem~\ref{thm:first approach}.
Here, we partition each matrix in the library by rows and let each sub-matrix be cached by a set of users. 

\begin{example}
\label{ex:row example}
\rm
Consider the $(\Ksf,\Nsf,\asf)=(4,20,1/2)$ shared-link cache-aided matrix multiplication  retrieval problem, with cache size $\Msf=10$. 
We use the cache replication strategy   in~\cite{finiteanalysis}. More precisely, we divide the $4$ users into $\ell \in [4]$ groups and let the users in the same group cache the same content. 
 
{\it\bf Case $\ell=4$.} 
First we consider the case $\ell=4$, in which case the  cache replication strategy   in~\cite{finiteanalysis} is the same as the MAN cache placement strategy in~\cite{dvbt2fundamental}. By computing $t_{4}=  \left\lfloor\frac{ 4 \Msf }{\Nsf} \right\rfloor=2$, we  partition  each matrix $\mathbf{W}_i$ where $i\in [20]$ into $\binom{\ell}{t_{\ell}}=6$ sub-matrices as follows (the dimension of a matrix is shown in the subscript of its parenthesis)
\begin{align}
(\mathbf{W}_i )_{\ssf \times \rsf} = \left[\begin{array}{c}
(\mathbf{W}_{i,\{1,2\}})_{\ssf/6 \times \rsf} \\ \hdashline
(\mathbf{W}_{i,\{1,3\}})_{\ssf/6 \times \rsf} \\ \hdashline
(\mathbf{W}_{i,\{1,4\}})_{\ssf/6 \times \rsf} \\ \hdashline
(\mathbf{W}_{i,\{2,3\}})_{\ssf/6 \times \rsf} \\ \hdashline
(\mathbf{W}_{i,\{2,4\}})_{\ssf/6 \times \rsf} \\ \hdashline
(\mathbf{W}_{i,\{3,4\}})_{\ssf/6 \times \rsf} 
\end{array}
\right].\label{eq:row Wi example}
\end{align}
Each sub-matrix $\mathbf{W}_{i, \Tc}$ where $\Tc \subseteq [4]$ and $|\Tc|=2$, is cached by users in $\Tc$. Thus, each user caches 
$
20 \times 3 \times \frac{\ssf \rsf }{6}  = 10 \ssf\rsf=\Msf \ssf \rsf
$
symbols in total, thus satisfying the cache size constraint.

Assume that 
\begin{align}
[\dv_1;\dv_2;\cdots;\dv_{4}]= [1,2;3,4;5,6;7,8].\label{eq:demand in the example}
\end{align}
The matrix product demanded by user~$1$ is 
\begin{align}
\mathbf{W}^{\text{\rm T}}_1  \mathbf{W}_2 &= \sum_{\Tc \subseteq [4]: |\Tc|=2} \mathbf{W}^{\text{\rm T}}_{1,\Tc} \mathbf{W}_{2,\Tc} \nonumber\\
&=\sum_{\Tc^{\prime} \subseteq [4]: |\Tc^{\prime}|=2, 1 \in \Tc^{\prime}} \mathbf{W}^{\text{\rm T}}_{1,\Tc^{\prime}} \mathbf{W}_{2,\Tc^{\prime}} + \sum_{\Tc \subseteq [4]: |\Tc|=2, 1 \notin \Tc} \mathbf{W}^{\text{\rm T}}_{1,\Tc} \mathbf{W}_{2,\Tc}.  \label{eq:ex ell 4 two term}
\end{align}
Note that the first term on the RHS of~\eqref{eq:ex ell 4 two term} is known by user~$1$ from its cache. Thus user~$1$ only needs to recover the second term. For each $\Tc \subseteq [4]$ where  $|\Tc|=2$ and $1 \notin \Tc$,  
$\mathbf{W}^{\text{\rm T}}_{1,\Tc} \mathbf{W}_{2,\Tc}$ is cached by the users in  $\Tc$. In addition, $\mathbf{W}^{\text{\rm T}}_{1,\Tc} \mathbf{W}_{2,\Tc}$ can be encoded into $ P(\mathbf{W}^{\text{\rm T}}_{1,\Tc}, \mathbf{W}_{2,\Tc})$  of size  $f(\rsf, \ssf/6,\rsf)$ symbols. 
Since $\asf=\frac{\rsf}{\ssf}=\frac{1}{2}$, we have 
\begin{align}
&f\left(\rsf, \frac{\ssf}{6},\rsf \right)=f\left( \frac{\ssf}{2},  \frac{\ssf}{6}, \frac{\ssf}{2} \right)   
 =\frac{\ssf}{6} (\frac{\ssf}{2}+\frac{\ssf}{2})- \left(\frac{\ssf}{6}\right)^2 
 =\frac{5 \ssf^2}{36}.
\end{align}
We will let user~$1$ recover $P(\mathbf{W}^{\text{\rm T}}_{1,\Tc}, \mathbf{W}_{2,\Tc})$ during the delivery phase.

After generating the coded symbols for each user, 
for each set  $\Sc \subseteq [\Ksf]$ where $|\Sc|=t_{4}+1=3$, the server broadcasts
\begin{align}
\sum_{k\in \Sc} P(\mathbf{W}^{\text{\rm T}}_{d_{k,1},\Sc \setminus \{k\}}, \mathbf{W}_{d_{k,2},\Sc\setminus \{k\}}).\label{eq:general row multicast}
\end{align}
 Each user~$k\in \Sc$  knows all the coded symbols in the sum~\eqref{eq:general row multicast} from its cache except $P(\mathbf{W}^{\text{\rm T}}_{d_{k,1},\Sc \setminus \{k\}}, \mathbf{W}_{d_{k,2},\Sc\setminus \{k\}})$, such that it can recover $P(\mathbf{W}^{\text{\rm T}}_{d_{k,1},\Sc \setminus \{k\}}, \mathbf{W}_{d_{k,2},\Sc\setminus \{k\}})$ and then recover $\mathbf{W}^{\text{\rm T}}_{d_{k,1},\Sc \setminus \{k\}}  \mathbf{W}_{d_{k,2},\Sc\setminus \{k\}}$. 
For example, for $\Sc=\{1,2,3\}$, the server broadcasts
 \begin{align}
  & P(\mathbf{W}^{\text{\rm T}}_{1,\{2,3\}}, \mathbf{W}_{2, \{2,3\}}) + P(\mathbf{W}^{\text{\rm T}}_{3,\{1,3\}}, \mathbf{W}_{4, \{1,3\}}) +P(\mathbf{W}^{\text{\rm T}}_{5,\{1,2\}}, \mathbf{W}_{6, \{1,2\}}),\label{eq:rwo 123}
 \end{align}
 and similarly for the remaining multicast messages. 
 Hence, the server broadcasts $4 f\left(\rsf, \frac{\ssf}{6},\rsf \right) =\frac{5 \ssf^2}{9}$ symbols in total, thus the achieved load is 
\begin{align}
 \frac{5 \ssf^2}{9 f(\rsf,\ssf,\rsf)}=  \frac{5 \ssf^2}{9 f(\ssf/2,\ssf,\ssf/2)}=\frac{20}{9}.
\end{align}
 
{\it\bf Case $\ell=2$.}
Then, we consider the case $\ell=2$. By computing  $t_{2}=  \left\lfloor\frac{ 2 \Msf }{\Nsf} \right\rfloor =1$, we  partition  each matrix $\mathbf{W}_i$ where $i\in [20]$ into $\binom{\ell}{t_{\ell}}=2$ sub-matrices as  
\begin{align}
( \mathbf{W}_i )_{\ssf \times \rsf} = \left[\begin{array}{c}
(\mathbf{W}_{i,\{1\}})_{\ssf/2 \times \rsf}  \\ \hdashline
(\mathbf{W}_{i,\{2\}})_{\ssf/2 \times \rsf} 
\end{array}
\right].\label{eq:row Wi example ell 2}
\end{align}
We let users~$1$ and $3$ cache $\mathbf{W}_{i,\{1\}}$, and let users~$2$ and $4$ cache $\mathbf{W}_{i,\{2\}}$.  
In other words, we divide the users into two placement groups, where the first group contains users $1$ and $3$, and the second group contains users $2$ and $4$. The users in the same group have the same cache content.
So each user caches 
$
20 \times   \frac{\ssf \rsf }{2}  = 10 \ssf\rsf=\Msf \ssf \rsf
$
symbols, satisfying the cache size constraint.

During the delivery phase, we assume that the users' demands are given as in~\eqref{eq:demand in the example}. 
The matrix product demanded by user~$1$ is 
\begin{align}
\mathbf{W}^{\text{\rm T}}_1  \mathbf{W}_2 &=  \mathbf{W}^{\text{\rm T}}_{1,\{1\}} \mathbf{W}_{2,\{1\}}+ \mathbf{W}^{\text{\rm T}}_{1,\{2\}} \mathbf{W}_{2,\{2\}},
\label{eq:ex ell 2 two term}
\end{align}
for which user~$1$ only needs to recover $\mathbf{W}^{\text{\rm T}}_{1,\{2\}} \mathbf{W}_{2,\{2\}}$.  
In addition, $\mathbf{W}^{\text{\rm T}}_{1,\{2\}} \mathbf{W}_{2,\{2\}} $ can be encoded into $ P(\mathbf{W}^{\text{\rm T}}_{1,\{2\}}, \mathbf{W}_{2,\{2\}})$  of size  $f\left(\rsf, \frac{\ssf}{2},\rsf \right)= \frac{\ssf^2}{4}$ symbols. 

After generating the coded symbols for each user, we divide the users into two transmission groups. 
In the first transmission group, we let the server satisfy the demands of users~$1$ and $2$ by broadcasting
 \begin{align}
   & P(\mathbf{W}^{\text{\rm T}}_{1,\{2\}}, \mathbf{W}_{2, \{2\}}) + P(\mathbf{W}^{\text{\rm T}}_{3,\{1\}}, \mathbf{W}_{4, \{1\}})  .\label{eq:rwo 12}
 \end{align}
In the transmission second group, we let the server satisfy the demands of users~$3$ and $4$ by broadcasting
 \begin{align}
   & P(\mathbf{W}^{\text{\rm T}}_{5,\{2\}}, \mathbf{W}_{6, \{2\}}) + P(\mathbf{W}^{\text{\rm T}}_{7,\{1\}}, \mathbf{W}_{8, \{1\}})  .\label{eq:rwo 34}
 \end{align}
 Hence, the server broadcasts $2 f\left(\rsf, \frac{\ssf}{2},\rsf \right) =\frac{  \ssf^2}{2}$ symbols in total, thus the achieved load is 
 $$
 \frac{  \ssf^2}{2 f(\rsf,\ssf,\rsf)}=  \frac{  \ssf^2}{ 2 f(\ssf/2,\ssf,\ssf/2)}=2.
 $$
 
{\it\bf Case $\ell=1$.} 
Similarly, when $\ell=1$  (i.e., one single placement group)  the achieved load is $4$.

{\it\bf Case $\ell=3$.}
When $\ell=3$ the achieved load is $\frac{40}{9}$ (i.e., three placement groups).

{\it\bf All Cases Together.}
Hence, the minimum load achieved by the proposed row-partition scheme is $2$ with $\ell=2$, which is less than $64/21$, $3$, and $8/3$ achieved by the structure-agnostic scheme in Theorem~\ref{thm:load of agnostic}  and the  two baseline  structure-aware  schemes in Theorem~\ref{thm:base}, respectively.
 \hfill $\square$ 
\end{example}

\begin{rem}[Row-partition: $\ell=4$ v.s. $\ell=2$]
\label{rem:ell4vs2}
\rm 
In Example~\ref{ex:row example}, 
when $\ell=4$, each transmitted packet is a sum of $t_{4}+1=3$ coded symbols, while when $\ell=2$, it is a sum of $t_{2}+1=2$ coded symbols. However, the latter attains the lowest load. This is because when $\ell=4$, in order to recover $\sum_{\Tc \subseteq [4]: |\Tc|=t_{4}=2, 1 \notin \Tc} \mathbf{W}^{\text{\rm T}}_{1,\Tc} \mathbf{W}_{2,\Tc}$ in~\eqref{eq:ex ell 4 two term}, we let user~$1$ recover each term in this sum, which increases the communication load. However, when $\ell=2$, there is one set $\Tc \subseteq [2]$ where $|\Tc|=t_{2}=1$ and $1 \notin \Tc$, and this set is $\Tc=\{2\}$; thus we directly let user~$1$ recover $\mathbf{W}^{\text{\rm T}}_{1,\{2\}} \mathbf{W}_{2,\{2\}}$ in~\eqref{eq:ex ell 2 two term}.

In other words, as already mentioned, the proposed row-partition scheme uses the cache replication placement in~\cite{finiteanalysis}, which was proposed for the MAN shared-link caching problem  for single file retrieval  in order to reduce the sub-packetization at the expense of a higher load compared to the MAN scheme.  However, in our row-partition approach for the considered cache-aided matrix multiplication  retrieval  problem, such a placement can simultaneously reduce the sub-packetization level  and the load compared to the MAN cache placement. 
\hfill $\square$ 
\end{rem}

We now generalize  the proposed row-partition scheme in Example~\ref{ex:row example}.
We focus on each $\ell \in [\Ksf]$. 

 \paragraph*{{ Placement phase}}
 We first compute   $ t_{\ell}=\left\lfloor \frac{ \ell \Msf}{\Nsf} \right\rfloor$ and $\alpha_{\ell}= t_{\ell}+1 -\frac{\ell \Msf}{\Nsf}$ as defined in~\eqref{eq:def of t1} and~\eqref{eq:def of alpha}, respectively.  
 Among all the $\ssf$ rows of each matrix in the library, there are $\alpha_{\ell} \ssf$ rows  cached by $t_{\ell}$ users, and  $(1-\alpha_{\ell}) \ssf$ rows   cached by $t_{\ell}+1$ users, such that the average number of users caching each row is $ \frac{ \ell \Msf}{\Nsf} $.
 More precisely, the first $\alpha_{\ell} \ssf$ rows of $\mathbf{W}_i$ where $i\in [\Nsf]$ are partitioned into $\binom{\ell}{t_{\ell}}$ sub-matrices, each of which is denoted by $\mathbf{W}_{i,\Tc_1}$ where $\Tc_1 \subseteq [\ell]$ and $|\Tc_1|=t_{\ell}$. $\mathbf{W}_{i,\Tc_1}$ has dimension $\frac{\alpha_{\ell} \ssf}{\binom{\ell}{t_{\ell}}} \times \rsf$.
 The remaining  $(1-\alpha_{\ell}) \ssf$ rows of $\mathbf{W}_i$ are partitioned into $\binom{\ell}{t_{\ell}+1}$ sub-matrices, each of which is denoted by $\mathbf{W}_{i,\Tc_2}$ where $\Tc_2 \subseteq [\ell]$ and $|\Tc_2|=t_{\ell}+1$. $\mathbf{W}_{i,\Tc_2}$ has dimension $\frac{(1-\alpha_{\ell}) \ssf}{\binom{\ell}{t_{\ell}+1}} \times \rsf$.
Each user~$k\in [\Ksf]$ caches $\mathbf{W}_{i,\Tc}$ where $i\in [\Nsf]$, $\Tc\subseteq [\ell]$, $|\Tc| \in \{ t_{\ell}, t_{\ell}+1\}$, and $\text{Mod}(k,\ell)\in \Tc$.\footnote{\label{foot:mod} $\text{Mod}(a,b)$  represents the modulo operation on $a$ with  integer quotient $b$. In this paper, if $b$ divides $a$, we let $\text{Mod}(a,b)=b$. } Hence, user~$k$ caches (recall that $\Msf=\frac{t_{\ell}+1-\alpha_{\ell}}{\ell}\Nsf$)
\begin{subequations}
\begin{align}
&\Nsf \left( \binom{\ell-1}{t_{\ell-1}}\frac{\alpha_{\ell} \ssf}{\binom{\ell}{t_{\ell}}} \cdot \rsf
+ \binom{\ell-1}{t_{\ell }} \frac{(1-\alpha_{\ell}) \ssf}{\binom{\ell}{t_{\ell}+1}} \cdot \rsf \right) \nonumber\\
&= \Nsf \ssf \rsf \left(  \frac{t_{\ell}}{\ell}\alpha_{\ell}+\frac{t_{\ell}+1}{\ell}(1-\alpha_{\ell}) \right)\\
&=\Nsf \ssf\rsf \frac{t_{\ell}+1-\alpha_{\ell}}{\ell}=\Msf \ssf \rsf \text{ symbols},
\end{align}
\end{subequations}
satisfying the cache size constraint.   

Note that  if $\text{Mod}(k_1,\ell)=\text{Mod}(k_2,\ell)$ where $k_1 , k_2 \in [\Ksf]$, users $k_1$ and $k_2$ have the same cache content.

  \paragraph*{{ Delivery phase}}
 For each $\ell \in [\Ksf]$, we define    
\begin{align}
 \Nc_{\ell}:=\big\{\Tc \subseteq [\ell]: |\Tc| \in \{t_{\ell}, t_{\ell}+1\}  \big\}, \label{eq:Nell}
\end{align} 
 and sort the sets in $ \Nc_{\ell}$   in a lexicographic order. $\Nc_{\ell}(j)$ represents the $j^{\text{th}}$ set in $\Nc_{\ell}$, where $j\in \left[ \binom{\ell+1}{t_{\ell}+1}\right]$.\footnote{\label{foot:pascal}From the Pascal's Triangle, we have $\binom{\ell}{t_{\ell}}+\binom{\ell}{t_{\ell}+1}=\binom{\ell+1}{t_{\ell}+1}$.}   
  
  We divide the users into $\left\lceil \frac{\Ksf}{\ell}\right\rceil $ groups. More precisely, we let   
  \begin{subequations}
\begin{align}
&\Gc_{i}=[(i-1)\ell+1 : i \ell], \ \ \forall i\in \left[ \left\lceil \frac{\Ksf}{\ell}-1 \right\rceil \right];\\
&\Gc_{\left\lceil \frac{\Ksf}{\ell}\right\rceil}= \left[\ell \left\lceil \frac{\Ksf}{\ell}-1 \right\rceil+1  :\Ksf\right],
\end{align}
\end{subequations}
where the first $\left\lceil \frac{\Ksf}{\ell}-1 \right\rceil $ groups contains $\ell$ users with different caches, and the last group contains $\Ksf-\ell \left\lceil \frac{\Ksf}{\ell}-1 \right\rceil$ users with different caches.  
 
 Let us focus on the transmission for group $\Gc_i$ where $i\in \left[\left\lceil \frac{\Ksf}{\ell}\right\rceil \right]$. We sort the users in $\Gc_i$ in an increasing order and let $\Gc_i(j)$ be the $j^{\text{th}}$ user.\footnote{\label{foot:empty} If $j>|\Gc_i|$, we let   $\Gc_i(j)=\emptyset$.}
For each user~$k\in \Gc_i$,  its desired matrix product can be expressed as  
     \begin{subequations}
\begin{align}
&\mathbf{W}^{\text{\rm T}}_{d_{k,1}}  \mathbf{W}_{d_{k,2}} = 
\left[\begin{array}{c}
 \mathbf{W}^{\text{\rm T}}_{d_{k,1},\Nc_{\ell}(1)}   \\ \hdashline
\vdots \\ \hdashline
 \mathbf{W}^{\text{\rm T}}_{d_{k,1},\Nc_{\ell}\left( \binom{\ell+1}{t_{\ell}+1} \right)}     
\end{array}\right]
\
\left[\begin{array}{c:c:c}
 \mathbf{W}_{d_{k,2},\Nc_{\ell}(1)}   &
\cdots &
 \mathbf{W}_{d_{k,2},\Nc_{\ell}\left( \binom{\ell+1}{t_{\ell}+1} \right)}   
\end{array} \right] \\
&= 
\sum_{\Tc_1 \subseteq [\ell]: |\Tc_1|=t_{\ell}} \mathbf{W}^{\text{\rm T}}_{d_{k,1},\Tc_1} \mathbf{W}_{d_{k,2},\Tc_1} +\sum_{\Tc_2 \subseteq [\ell]: |\Tc_2|=t_{\ell}+1} \mathbf{W}^{\text{\rm T}}_{d_{k,1},\Tc_2} \mathbf{W}_{d_{k,2},\Tc_2}  \\
&=\sum_{\Tc^{\prime}_1 \subseteq [\ell]: |\Tc^{\prime}_1|=t_{\ell}, \text{Mod}(k,\ell) \in \Tc^{\prime}_1} \mathbf{W}^{\text{\rm T}}_{d_{k,1},\Tc^{\prime}_1} \mathbf{W}_{d_{k,2},\Tc^{\prime}_1} +
\sum_{\Tc_1 \subseteq [\ell]: |\Tc_1|=t_{\ell},\text{Mod}(k,\ell) \notin \Tc_1} \mathbf{W}^{\text{\rm T}}_{d_{k,1},\Tc_1} \mathbf{W}_{d_{k,2},\Tc_1} \nonumber\\
&+\sum_{\Tc^{\prime}_2 \subseteq [\ell]: |\Tc^{\prime}_2|=t_{\ell}+1,\text{Mod}(k,\ell) \in \Tc^{\prime}_2 } \mathbf{W}^{\text{\rm T}}_{d_{k,1},\Tc^{\prime}_2} \mathbf{W}_{d_{k,2},\Tc^{\prime}_2}
\sum_{\Tc_2 \subseteq [\ell]: |\Tc_2|=t_{\ell}+1,\text{Mod}(k,\ell) \notin \Tc_2} \mathbf{W}^{\text{\rm T}}_{d_{k,1},\Tc_2} \mathbf{W}_{d_{k,2},\Tc_2}.  \label{eq: ell 4 term}
\end{align}
      \end{subequations}
We note that the first and third term on the RHS of~\eqref{eq: ell 4 term} can be re-constructed by the cached content of user~$k$. Hence, user~$k$ only needs to recover the second and fourth terms in~\eqref{eq: ell 4 term} during the delivery phase.
For each  $\Tc_1 \subseteq [\ell]$ where $|\Tc_1|=t_{\ell}$ and $\text{Mod}(k,\ell) \notin \Tc_1$, we can encode $\mathbf{W}^{\text{\rm T}}_{d_{k,1},\Tc_1} \mathbf{W}_{d_{k,2},\Tc_1}$ into $P\left(\mathbf{W}^{\text{\rm T}}_{d_{k,1},\Tc_1} ,\mathbf{W}_{d_{k,2},\Tc_1} \right)$ of size  
\begin{align}
f\left(\rsf, \frac{\alpha_{\ell} \ssf}{\binom{\ell}{t_{\ell}}}, \rsf \right)= g\left(\frac{\rsf \binom{\ell}{t_{\ell}} }{ \alpha_{\ell} \ssf} ,\frac{\rsf \binom{\ell}{t_{\ell}} }{ \alpha_{\ell} \ssf} \right)  \left(\frac{\alpha_{\ell} \ssf}{\binom{\ell}{t_{\ell}}}\right)^2=g\left(\frac{\asf \binom{\ell}{t_{\ell}} }{ \alpha_{\ell} } ,\frac{\asf \binom{\ell}{t_{\ell}} }{ \alpha_{\ell}} \right)  \left(\frac{\alpha_{\ell} \ssf}{\binom{\ell}{t_{\ell}}}\right)^2 \text{ symbols},
\end{align}
where $\asf=\rsf/\ssf$, and $f(\cdot)$ and $g(\cdot)$ are defined in Section~\ref{sec:pre}. 
We will let user~$k$ recover $P\left(\mathbf{W}^{\text{\rm T}}_{d_{k,1},\Tc_1}, \mathbf{W}_{d_{k,2},\Tc_1} \right)$ during the delivery phase.  For each  $\Tc_2 \subseteq [\ell]$ where $|\Tc_2|=t_{\ell}+1$ and $\text{Mod}(k,\ell) \notin \Tc_2$, we can encode $\mathbf{W}^{\text{\rm T}}_{d_{k,1},\Tc_2} \mathbf{W}_{d_{k,2},\Tc_2}$ into $P\left(\mathbf{W}^{\text{\rm T}}_{d_{k,1},\Tc_2}, \mathbf{W}_{d_{k,2},\Tc_2} \right)$  of size  
     \begin{subequations}
\begin{align}
f\left(\rsf, \frac{(1-\alpha_{\ell}) \ssf}{\binom{\ell}{t_{\ell}+1}} , \rsf \right)&= g\left(\frac{\rsf \binom{\ell}{t_{\ell}+1} }{(1-\alpha_{\ell}) \ssf} ,\frac{\rsf \binom{\ell}{t_{\ell}+1} }{(1-\alpha_{\ell}) \ssf} \right)  \left(\frac{(1-\alpha_{\ell}) \ssf}{\binom{\ell}{t_{\ell}+1}} \right)^2
\\&=g\left(\frac{\asf \binom{\ell}{t_{\ell}+1} }{(1-\alpha_{\ell}) } ,\frac{\asf \binom{\ell}{t_{\ell}+1} }{(1-\alpha_{\ell}) } \right)  \left(\frac{(1-\alpha_{\ell}) \ssf}{\binom{\ell}{t_{\ell}+1}} \right)^2 \text{ symbols}.
\end{align}
     \end{subequations}
We will also let user~$k$ recover $P\left(\mathbf{W}^{\text{\rm T}}_{d_{k,1},\Tc_2} ,\mathbf{W}_{d_{k,2},\Tc_2} \right)$ during the delivery phase.

 After generating the desired coded symbols for all users in $\Gc_i$,
 for each set $\Sc_1 \subseteq [\ell]$ where $\Sc_1=t_{\ell}+1$, the server broadcasts
\begin{align}
X_{i, \Sc_1}:= \sum_{j \in \Sc_1} P\left(\mathbf{W}^{\text{\rm T}}_{d_{(i-1)\ell+j,1},\Sc_1 \setminus \{j\} }, \mathbf{W}_{d_{(i-1)\ell+j,2},\Sc_1 \setminus \{j\} } \right). \label{eq:X i S1}
\end{align} 
So user~$  (i-1)\ell+j $  can recover $P\left(\mathbf{W}^{\text{\rm T}}_{d_{(i-1)\ell+j,1},\Sc_1 \setminus \{j\} }, \mathbf{W}_{d_{(i-1)\ell+j,2},\Sc_1 \setminus \{j\} } \right)$
from $X_{i,\Sc_1}$, where $j \in \Sc_1$. Similarly, for each set   $\Sc_2 \subseteq [\ell]$ where $\Sc_2=t_{\ell}+2$, the server broadcasts
\begin{align}
X_{i, \Sc_2}:= \sum_{j \in \Sc_2} P\left(\mathbf{W}^{\text{\rm T}}_{d_{(i-1)\ell+j,1},\Sc_2 \setminus \{j\} }, \mathbf{W}_{d_{(i-1)\ell+j,2},\Sc_2 \setminus \{j\} } \right), \label{eq:X i S2}
\end{align} 
 such that user  $ (i-1)\ell+j $  can recover $P\left(\mathbf{W}^{\text{\rm T}}_{d_{(i-1)\ell+j,1},\Sc_2 \setminus \{j\} } ,\mathbf{W}_{d_{(i-1)\ell+j,2},\Sc_2 \setminus \{j\} } \right)$.
Hence, for the users in  $\Gc_i$, the total number of symbols transmitted by the server is 
$$
\binom{\ell}{t_{\ell}+1} g\left(\frac{\asf \binom{\ell}{t_{\ell}} }{ \alpha_{\ell} } ,\frac{\asf \binom{\ell}{t_{\ell}} }{ \alpha_{\ell}} \right)  \left(\frac{\alpha_{\ell} \ssf}{\binom{\ell}{t_{\ell}}}\right)^2 + \binom{\ell}{t_{\ell}+2}g\left(\frac{\asf \binom{\ell}{t_{\ell}+1} }{(1-\alpha_{\ell}) } ,\frac{\asf \binom{\ell}{t_{\ell}+1} }{(1-\alpha_{\ell}) } \right)  \left(\frac{(1-\alpha_{\ell}) \ssf}{\binom{\ell}{t_{\ell}+1}} \right)^2.
$$
 
 Considering all the $\left\lceil \frac{\Ksf}{\ell}\right\rceil$ transmission groups, the total load is 
 \begin{align}
 &\left\lceil \frac{\Ksf}{\ell}\right\rceil \frac{\binom{\ell}{t_{\ell}+1} g\left(\frac{\asf \binom{\ell}{t_{\ell}} }{ \alpha_{\ell} } ,\frac{\asf \binom{\ell}{t_{\ell}} }{ \alpha_{\ell}} \right)  \left(\frac{\alpha_{\ell} \ssf}{\binom{\ell}{t_{\ell}}}\right)^2 + \binom{\ell}{t_{\ell}+2}g\left(\frac{\asf \binom{\ell}{t_{\ell}+1} }{(1-\alpha_{\ell}) } ,\frac{\asf \binom{\ell}{t_{\ell}+1} }{(1-\alpha_{\ell}) } \right)  \left(\frac{(1-\alpha_{\ell}) \ssf}{\binom{\ell}{t_{\ell}+1}} \right)^2}{ f(\rsf,\ssf,\rsf)} \nonumber\\
 &= \left\lceil \frac{\Ksf}{\ell}\right\rceil \frac{\binom{\ell}{t_{\ell}+1} g\left(\frac{\asf \binom{\ell}{t_{\ell}} }{ \alpha_{\ell} } ,\frac{\asf \binom{\ell}{t_{\ell}} }{ \alpha_{\ell}} \right)  \left(\frac{\alpha_{\ell} }{\binom{\ell}{t_{\ell}}}\right)^2 + \binom{\ell}{t_{\ell}+2}g\left(\frac{\asf \binom{\ell}{t_{\ell}+1} }{(1-\alpha_{\ell}) } ,\frac{\asf \binom{\ell}{t_{\ell}+1} }{(1-\alpha_{\ell}) } \right)  \left(\frac{(1-\alpha_{\ell}) }{\binom{\ell}{t_{\ell}+1}} \right)^2}{ g(\asf,\asf) },
 \end{align}
 which coincides with~\eqref{eq:R_row}.

\begin{rem}[Row-partition scheme v.s. multi-request baseline scheme]
 \label{rem:proof cor formula 1}
 \rm
 If we let $\ell=\Ksf$, encode $\mathbf{W}^{\text{\rm T}}_{d_{k,1},\Tc_1} \mathbf{W}_{d_{k,2},\Tc_1}$ into the concatenation of all the symbols in $\mathbf{W}^{\text{\rm T}}_{d_{k,1},\Tc_1}$ and $ \mathbf{W}_{d_{k,2},\Tc_1}$ whose length   is strictly larger than the length of $P\left(\mathbf{W}^{\text{\rm T}}_{d_{k,1},\Tc_1} ,\mathbf{W}_{d_{k,2},\Tc_1} \right)$, and encode $\mathbf{W}^{\text{\rm T}}_{d_{k,1},\Tc_2} \mathbf{W}_{d_{k,2},\Tc_2}$ into the concatenation of all the symbols in $\mathbf{W}^{\text{\rm T}}_{d_{k,1},\Tc_2}$ and $ \mathbf{W}_{d_{k,2},\Tc_2}$  whose length     is  strictly larger than the length of $P\left(\mathbf{W}^{\text{\rm T}}_{d_{k,1},\Tc_2} ,\mathbf{W}_{d_{k,2},\Tc_2} \right)$,  then
it is equivalent to let each user recover the two library matrices of its desired matrix product; thus the proposed row-partition scheme becomes the multi-request baseline scheme. Therefore, the proposed row-partition scheme  is strictly better  than the multi-request baseline scheme when $\Msf<\Nsf$.
  \hfill $\square$ 
 \end{rem}
 
 \begin{rem}[Application of other shared-link coded caching schemes]
 \label{rem:all caching}
 \rm
Obviously, with the proposed row-partition strategy, we can apply any shared-link coded caching scheme with uncoded cache placement for the original MAN coded caching problem for  single file  retrieval to the considered cache-aided matrix multiplication  retrieval  problem. More precisely, for any existing scheme with uncoded cache placement for the  single file retrieval problem,  each file $W^{\prime}_i$ where $i\in [\Nsf]$ is divided into non-overlapping subfiles, $W^{\prime}_i=\{W^{\prime}_{i,\Tc}:\Tc \subseteq [\Ksf]\}$.
In the considered matrix multiplication  retrieval  problem, we can partition each  matrix $\mathbf{W}_i$ into $2^{\Ksf}$ sub-matrix by rows, each sub-matrix denoted by $\mathbf{W}_{i,\Tc}$ of  dimension $\frac{\ssf |W^{\prime}_{i,\Tc}|}{|W^{\prime}_{i}|} \times \rsf$. We then encode $\mathbf{W}^{\text{\rm T}}_{d_{k,1},\Tc} \mathbf{W}_{d_{k,2},\Tc}$ into $P\left(\mathbf{W}^{\text{\rm T}}_{d_{k,1},\Tc} \mathbf{W}_{d_{k,2},\Tc} \right)$ symbols. Finally, we use the delivery phase of this existing scheme to deliver $P\left(\mathbf{W}^{\text{\rm T}}_{d_{k,1},\Tc} \mathbf{W}_{d_{k,2},\Tc} \right)$ as delivering $W^{\prime}_{d_k,\Tc}$ in the original file retrieval problem, where $d_k$ represents the desired file of user~$k$.
  \hfill $\square$ 
 \end{rem}

\

\subsection{Column-partition Scheme: Proof of Theorem~\ref{thm:second approach}}
\label{sub:second approach}
We continue Example~\ref{ex:row example} to introduce the column-partition scheme in Theorem~\ref{thm:second approach}.
Here we partition each matrix in the library by columns and let each sub-matrix be cached by a set of users.

\begin{example}
\label{ex:column example}
\rm
Recall that we consider the $(\Ksf,\Nsf,\asf)=(4,20,1/2)$ shared-link cache-aided matrix multiplication  retrieval  problem with cache size $\Msf=10$.

 \paragraph*{{ Placement phase}}
We use the MAN cache placement in~\cite{dvbt2fundamental}.  
With $t_{\Ksf}=  \left\lfloor\frac{\Ksf \Msf }{\Nsf} \right\rfloor=2$, we  partition  each matrix $\mathbf{W}_i$ where $i\in [20]$ into $\binom{\Ksf}{t_{\Ksf}}=6$ sub-matrices as follows: 
\begin{align}
( \mathbf{W}_i )_{\ssf \times \rsf} = \left[\begin{array}{c:c:c:c:c:c}
(\mathbf{W}_{i,\{1,2\}})_{\ssf \times \frac{\rsf}{6}}   &
(\mathbf{W}_{i,\{1,3\}})_{\ssf \times \frac{\rsf}{6}}   &
(\mathbf{W}_{i,\{1,4\}})_{\ssf \times \frac{\rsf}{6}}   &
(\mathbf{W}_{i,\{2,3\}})_{\ssf \times \frac{\rsf}{6}}   &
(\mathbf{W}_{i,\{2,4\}})_{\ssf \times \frac{\rsf}{6}}   &
(\mathbf{W}_{i,\{3,4\}})_{\ssf \times \frac{\rsf}{6}} 
\end{array}
\right],
\label{eq:col Wi example}
\end{align}
where sub-matrix $\mathbf{W}_{i, \Tc},$ for $\Tc \subseteq [4]$ and $|\Tc|=2$, is cached by users in $\Tc$. 
Each user thus caches 
$
20 \times 3 \times  \frac{\ssf \rsf }{6}  = 10 \ssf\rsf=\Msf \ssf \rsf
$
symbols, satisfying the cache size constraint.

\paragraph*{{ Delivery phase}}
Assume that the users' demands are as in~\eqref{eq:demand in the example}. 
The matrix product  demanded by user~$1$ can be expressed as
\begin{align}
(\mathbf{W}^{\text{\rm T}}_{1 }  \mathbf{W}_{2 })_{\rsf \times \rsf} = \left[\begin{array}{c:c:c}
(\mathbf{W}^{\text{\rm T}}_{1,\{1,2\}} \mathbf{W}_{2,\{1,2\}})_{\frac{\rsf}{6} \times \frac{\rsf}{6}}   & \cdots  &(\mathbf{W}^{\text{\rm T}}_{1,\{1,2\}} \mathbf{W}_{2,\{3,4\}})_{\frac{\rsf}{6} \times \frac{\rsf}{6}} \\ \hdashline
(\mathbf{W}^{\text{\rm T}}_{1,\{1,3\}} \mathbf{W}_{2,\{1,2\}})_{\frac{\rsf}{6} \times \frac{\rsf}{6}}   & \cdots  &(\mathbf{W}^{\text{\rm T}}_{1,\{1,3\}} \mathbf{W}_{2,\{3,4\}})_{\frac{\rsf}{6} \times \frac{\rsf}{6}}\\ \hdashline
\vdots  & \ddots  &\vdots \\  \hdashline
(\mathbf{W}^{\text{\rm T}}_{1,\{3,4\}} \mathbf{W}_{2,\{1,2\}})_{\frac{\rsf}{6} \times \frac{\rsf}{6}}   & \cdots  &(\mathbf{W}^{\text{\rm T}}_{1,\{3,4\}} \mathbf{W}_{2,\{3,4\}})_{\frac{\rsf}{6} \times \frac{\rsf}{6}}
\end{array}
\right].\label{eq:col demand user 1 example}
\end{align}
 Each sub-matrix $\mathbf{W}^{\text{\rm T}}_{1,\Tc_1} \mathbf{W}_{2,\Tc_2}$ in~\eqref{eq:col demand user 1 example} where $\Tc_1, \Tc_2\subseteq [4]$ and $|\Tc_1|=|\Tc_2|=2$,
 is then encoded into $P(\mathbf{W}^{\text{\rm T}}_{1,\Tc_1}, \mathbf{W}_{2,\Tc_2})$ of
 $
 f\left(\frac{\rsf}{6}, \ssf, \frac{\rsf}{6} \right)=  f\left(\frac{\ssf}{12}, \ssf, \frac{\ssf}{12} \right)=\frac{\ssf^2}{144}
 $
  symbols. Note that $P(\mathbf{W}^{\text{\rm T}}_{1,\Tc_1}, \mathbf{W}_{2,\Tc_2})$ can be directly re-constructed by each user in $\Tc_1 \cap \Tc_2$ from their cached content. Hence, during the delivery phase user~$1$ needs to recover $P(\mathbf{W}^{\text{\rm T}}_{1,\Tc_1}, \mathbf{W}_{2,\Tc_2})$ where $1 \notin (\Tc_1 \cap \Tc_2)$. We divide the coded symbols  desired by user~$1$ into groups, such that $F_{1,\Vc}$ represents the set of coded symbols desired by user~$1$ and uniquely known by users in $\Vc$. More precisely, we have
  \begin{subequations}
  \begin{align}
     F_{1, \emptyset} &=\Big\{P(\mathbf{W}^{\text{\rm T}}_{1,\{1,2 \}}, \mathbf{W}_{2,\{3,4\}}), \  P(\mathbf{W}^{\text{\rm T}}_{1,\{1,3 \}}, \mathbf{W}_{2,\{2,4\}}), \ P(\mathbf{W}^{\text{\rm T}}_{1,\{1,4 \}}, \mathbf{W}_{2,\{2,3\}}),\nonumber \\ & P(\mathbf{W}^{\text{\rm T}}_{1,\{2,3 \}}, \mathbf{W}_{2,\{1,4\}}), \ P(\mathbf{W}^{\text{\rm T}}_{1,\{2,4 \}}, \mathbf{W}_{2,\{1,3\}}),\  P(\mathbf{W}^{\text{\rm T}}_{1,\{3,4 \}}, \mathbf{W}_{2,\{1,2\}}) \Big\}; \label{eq: F_{1,0}}\\
   F_{1, \{2\}} &=\Big\{P(\mathbf{W}^{\text{\rm T}}_{1,\{1,2 \}}, \mathbf{W}_{2,\{2,3\}}), \  P(\mathbf{W}^{\text{\rm T}}_{1,\{1,2 \}}, \mathbf{W}_{2,\{2,4\}}), \ P(\mathbf{W}^{\text{\rm T}}_{1,\{2,3 \}}, \mathbf{W}_{2,\{1,2\}}),\nonumber \\ & P(\mathbf{W}^{\text{\rm T}}_{1,\{2,3 \}}, \mathbf{W}_{2,\{2,4\}}), \ P(\mathbf{W}^{\text{\rm T}}_{1,\{2,4 \}}, \mathbf{W}_{2,\{1,2\}}),\  P(\mathbf{W}^{\text{\rm T}}_{1,\{2,4 \}}, \mathbf{W}_{2,\{2,3\}}) \Big\} ;\label{eq: F_{1,2}}\\
   F_{1, \{3\}} &=\Big\{P(\mathbf{W}^{\text{\rm T}}_{1,\{1,3 \}}, \mathbf{W}_{2,\{2,3\}}), \  P(\mathbf{W}^{\text{\rm T}}_{1,\{1,3 \}}, \mathbf{W}_{2,\{3,4\}}), \ P(\mathbf{W}^{\text{\rm T}}_{1,\{2,3 \}}, \mathbf{W}_{2,\{1,3\}}),\nonumber \\ & P(\mathbf{W}^{\text{\rm T}}_{1,\{2,3 \}}, \mathbf{W}_{2,\{3,4\}}), \ P(\mathbf{W}^{\text{\rm T}}_{1,\{3,4 \}}, \mathbf{W}_{2,\{1,3\}}),\  P(\mathbf{W}^{\text{\rm T}}_{1,\{3,4 \}}, \mathbf{W}_{2,\{2,3\}}) \Big\}; \label{eq: F_{1,3}}\\
     F_{1, \{4\}} &=\Big\{P(\mathbf{W}^{\text{\rm T}}_{1,\{1,4 \}}, \mathbf{W}_{2,\{2,4\}}), \  P(\mathbf{W}^{\text{\rm T}}_{1,\{1,4 \}}, \mathbf{W}_{2,\{3,4\}}), \ P(\mathbf{W}^{\text{\rm T}}_{1,\{2,4 \}}, \mathbf{W}_{2,\{1,4\}}),\nonumber \\ & P(\mathbf{W}^{\text{\rm T}}_{1,\{2,4 \}}, \mathbf{W}_{2,\{3,4\}}), \ P(\mathbf{W}^{\text{\rm T}}_{1,\{3,4 \}}, \mathbf{W}_{2,\{1,4\}}),\  P(\mathbf{W}^{\text{\rm T}}_{1,\{3,4 \}}, \mathbf{W}_{2,\{2,4\}}) \Big\} ;\label{eq: F_{1,4}}\\
         F_{1, \{2,3\}} &=\Big\{P(\mathbf{W}^{\text{\rm T}}_{1,\{2,3 \}}, \mathbf{W}_{2,\{2,3\}})  \Big\} ;\label{eq: F_{1,23}}\\
               F_{1, \{2,4\}} &=\Big\{P(\mathbf{W}^{\text{\rm T}}_{1,\{2,4 \}}, \mathbf{W}_{2,\{2,4\}})  \Big\}; \label{eq: F_{1,24}}\\
                     F_{1, \{3,4\}} &=\Big\{P(\mathbf{W}^{\text{\rm T}}_{1,\{3,4 \}}, \mathbf{W}_{2,\{3,4\}})  \Big\}. \label{eq: F_{1,34}} 
  \end{align}
  \label{eq:ex1 col user 1 demand}
 \end{subequations}
 From~\eqref{eq:ex1 col user 1 demand}, 
%
and similarly for the other users, we have 
\begin{subequations}
  \begin{align}
  |F_{2, \emptyset} |&=|F_{3, \emptyset} |=|F_{4, \emptyset} |=\frac{\ssf^2}{24}; \label{eq:|F0| user 234}\\
  |F_{2,\{1\}}|&=|F_{2,\{3\}}|=|F_{2,\{4\}}|= |F_{3,\{1\}}|=|F_{3,\{2\}}| \nonumber\\& =|F_{3,\{4\}}|=|F_{4,\{1\}}|=|F_{4,\{2\}}|=|F_{4,\{3\}}|
 =\frac{\ssf^2}{24}; \label{eq:|F1| user 234}\\
  |F_{2,\{1,3\}}|& =|F_{2,\{1,4\}}|=|F_{2,\{3,4\}}|= |F_{3,\{1,2\}}|=|F_{3,\{1,4\}}|=|F_{3,\{2,4\}}| \nonumber\\&=|F_{4,\{1,2\}}|=|F_{4,\{1,3\}}|=|F_{4,\{2,3\}}|
 =\frac{\ssf^2}{144}. 
 \label{eq:|F2| user 234} 
  \end{align}
\end{subequations}
     
     Next we divide the transmission into three rounds. 
     In the first round, the server broadcasts
     \begin{align}
     F_{1,\emptyset}, \ \  F_{2,\emptyset}, \ \ F_{3,\emptyset}, \ \  F_{4, \emptyset},
     \end{align}
     for a total of $\frac{4\ssf^2}{24}=\frac{\ssf^2}{6}$ symbols.
     In the second round, the server broadcasts
     \begin{align}
   &  F_{1,\{2\}}+ F_{2,\{1\}}, \  \ F_{1,\{3\}}+ F_{3,\{1\}}, \  \ F_{1,\{4\}}+ F_{4,\{1\}}, \nonumber\\
   & F_{2,\{3\}}+ F_{3,\{2\}}, \  \ F_{2,\{4\}}+ F_{4,\{2\}}, \  \ F_{3,\{4\}}+ F_{4,\{3\}},
     \end{align}
     for a total of $\frac{6\ssf^2}{24}=\frac{\ssf^2}{4}$ symbols.
     In the third round, the server broadcasts
     \begin{align}
   & F_{1,\{2,3\}}+ F_{2,\{1,3\}}+F_{3,\{1,2\}}, \  \ F_{1,\{2,4\}}+ F_{2,\{1,4\}}+F_{4,\{1,2\}}, \nonumber\\
   & F_{1,\{3,4\}}+ F_{3,\{1,4\}}+F_{4,\{1,3\}}, \  \ F_{2,\{3,4\}}+ F_{3,\{2,4\}}+F_{4,\{2,3\}},
     \end{align}
     for a total of $\frac{4\ssf^2}{144}=\frac{\ssf^2}{36}$ symbols.      
       Hence,  the achieved load is 
       $$
       \frac{\frac{\ssf^2}{6}+\frac{\ssf^2}{4}+\frac{\ssf^2}{36}}{f(\rsf,\ssf,\rsf)}=   \frac{\frac{\ssf^2}{6}+\frac{\ssf^2}{4}+\frac{\ssf^2}{36}}{f(\ssf/2,\ssf,\ssf/2)}= \frac{16}{9},
       $$
       which is less than all  other  schemes.
   \hfill $\square$ 
\end{example}

\begin{rem}[Row-partition with $\ell= \Ksf $ v.s. column-partition]
\label{rem:row vs column}
\rm 
We now compare the row-partition scheme with   $\ell= \Ksf =4$ and the column-partition scheme through the above example. 
In both schemes, each sub-matrix in the library matrices is cached by $t_{4}=2$ users.
The main advantage of the row-partition scheme with $\ell=4$ is that each transmitted packet is a sum of $t_{4}+1=3$ coded symbols, while  most packets transmitted by the 
column-partition scheme are the sums of $t_{4}=2$ coded symbols. However,   each element in the desired matrix product by each user is a sum of some products of the elements in the library matrices. Instead of letting the user recover each individual product in the sum as in the row-partition scheme (e.g., we let user~$1$ recover each individual product in the sum~\eqref{eq:ex ell 4 two term}), the column-partition scheme directly lets the user recover this sum (e.g., we let user~$1$ recover each term in the product matrix~\eqref{eq:col demand user 1 example}).

To conclude, as mentioned already in Section~\ref{sub:highlevel}, 
the main advantage of the row-partition scheme is to fully leverage the multicast opportunities, while the main advantage of the column-partition scheme is to let each user directly recover each element in the product.
 \hfill $\square$ 
\end{rem}

We then generalize the column-partition scheme in Example~\ref{ex:column example}.

\subsubsection{$\asf \leq 1$}
\label{subsub:case a <1}
Let us first consider the case where $\asf \leq 1$ (i.e., $\rsf \leq  \ssf$).
 \paragraph*{{ Placement phase}}
Let $ t_{\Ksf}=\left\lfloor \frac{ \Ksf \Msf}{\Nsf} \right\rfloor$ and $\alpha_{\Ksf}= t_{\Ksf}+1 -\frac{\Ksf \Msf}{\Nsf}$. Among all the $\rsf$ columns of each matrix in the library, there $\alpha_{\Ksf} \rsf$ columns 
cached by $t_{\Ksf}$ users, and  $(1-\alpha_{\Ksf}) \rsf$ columns 
cached by $t_{\Ksf}+1$ users, such that the average number of users caching each column is $ \frac{ \Ksf\Msf}{\Nsf} $.
   More precisely, the first $\alpha_{\Ksf} \rsf$ columns of $\mathbf{W}_i$ where $i\in [\Nsf]$ are partitioned into $\binom{\Ksf}{t_{\Ksf}}$ sub-matrices, each of which is denoted by $\mathbf{W}_{i,\Tc_1}$ where $\Tc_1 \subseteq [\Ksf]$ and $|\Tc_1|=t_{\Ksf}$. $\mathbf{W}_{i,\Tc_1}$ has dimension $  \ssf \times \frac{\alpha_{\Ksf} \rsf}{\binom{\Ksf}{t_{\Ksf}}}$.
 The remaining  $(1-\alpha_{\Ksf}) \rsf$ columns of $\mathbf{W}_i$ are partitioned into $\binom{\Ksf}{t_{\Ksf}+1}$ sub-matrices, each of which is denoted by $\mathbf{W}_{i,\Tc_2}$ where $\Tc_2 \subseteq [\Ksf]$ and $|\Tc_2|=t_{\Ksf}+1$. $\mathbf{W}_{i,\Tc_2}$ has dimension $  \ssf \times \frac{(1-\alpha_{\Ksf}) \rsf}{\binom{\Ksf}{t_{\Ksf}+1}} $.
  
 Each user~$k\in [\Ksf]$ caches $\mathbf{W}_{i,\Tc}$ where $i\in [\Nsf]$, $\Tc\subseteq [\Ksf]$, $|\Tc| \in \{ t_{\Ksf}, t_{\Ksf}+1\}$, and $k\in \Tc$. Hence, user~$k\in [\Ksf]$ caches
 \begin{subequations}
\begin{align}
&\Nsf \left(  \binom{\Ksf-1}{t_{\Ksf-1}} \ssf \cdot \frac{\alpha_{\Ksf} \rsf}{\binom{\Ksf}{t_{\Ksf}}} 
 + \binom{\Ksf-1}{t_{\Ksf }} \ssf \cdot \frac{(1-\alpha_{\Ksf}) \rsf}{\binom{\Ksf}{t_{\Ksf}+1}}  \right) \nonumber\\
&= \Nsf \ssf \rsf \left(  \frac{t_{\Ksf}}{\Ksf}\alpha_{\Ksf}+\frac{t_{\Ksf}+1}{\Ksf}(1-\alpha_{\Ksf}) \right)\\
&=\Nsf \ssf\rsf \frac{t_{\Ksf}+1-\alpha_{\Ksf}}{\Ksf}=\Msf \ssf \rsf \ \text{ symbols},
\end{align}
 \end{subequations}
thus satisfying the cache size constraint.

   \paragraph*{{ Delivery phase}}
 Recall from~\eqref{eq:Nell} that       
 $
 \Nc_{\Ksf}:=\big\{\Tc \subseteq [\Ksf]: |\Tc| \in \{t_{\Ksf}, t_{\Ksf}+1\}  \big\},
 $ 
 where $|\Nc_{\Ksf}|= \binom{\Ksf+1}{t_{\Ksf}+1}$.
Let $\Nc_{\Ksf}(j)$ represents the $j^{\text{th}}$ set in $\Nc_{\Ksf}$, where $j\in \left[ \binom{\Ksf+1}{t_{\Ksf}+1}\right]$.
 
 The matrix product desired by user~$k\in[\Ksf]$ can be expressed as
\begin{align}
 \mathbf{W}^{\text{\rm T}}_{d_{k,1} }  \mathbf{W}_{d_{k,2} }  = \left[\begin{array}{c:c:c}
 \mathbf{W}^{\text{\rm T}}_{d_{k,1},\Nc_{\Ksf}(1)} \mathbf{W}_{d_{k,2},\Nc_{\Ksf}(1)}   & \cdots  & \mathbf{W}^{\text{\rm T}}_{d_{k,1},\Nc_{\Ksf}(1)} \mathbf{W}_{d_{k,2},\Nc_{\Ksf}\left( \binom{\Ksf+1}{t_{\Ksf}+1}\right)}  \\ \hdashline
 \vdots   & \ddots  &  \vdots \\ \hdashline
 \mathbf{W}^{\text{\rm T}}_{d_{k,1},\Nc_{\Ksf}\left( \binom{\Ksf+1}{t_{\Ksf}+1}\right)} \mathbf{W}_{d_{k,2},\Nc_{\Ksf}(1)} & \cdots  & \mathbf{W}^{\text{\rm T}}_{d_{k,1},\Nc_{\Ksf}\left( \binom{\Ksf+1}{t_{\Ksf}+1}\right)} \mathbf{W}_{d_{k,2},\Nc_{\Ksf}\left( \binom{\Ksf+1}{t_{\Ksf}+1}\right)}  
\end{array}
\right].\label{eq:col demand user k}
\end{align}
For any pair $(j_1,j_2)$ where $j_1, j_2 \in \left[ \binom{\Ksf+1}{t_{\Ksf}+1}\right]$, 
\begin{itemize}

\item if $k \in \Nc_{\Ksf}(j_1) \cap \Nc_{\Ksf}(j_2)$, 
 $ \mathbf{W}^{\text{\rm T}}_{d_{k,1},\Nc_{\Ksf}(j_1)} \mathbf{W}_{d_{k,2},\Nc_{\Ksf}(j_2)} $ can be reconstructed by user~$k$ from its cached content;

\item otherwise, we encode  $ \mathbf{W}^{\text{\rm T}}_{d_{k,1},\Nc_{\Ksf}(j_1)} \mathbf{W}_{d_{k,2},\Nc_{\Ksf}(j_2)} $  into $P\left( \mathbf{W}^{\text{\rm T}}_{d_{k,1},\Nc_{\Ksf}(j_1)}, \mathbf{W}_{d_{k,2},\Nc_{\Ksf}(j_2)} \right)$. We then add $P\left( \mathbf{W}^{\text{\rm T}}_{d_{k,1},\Nc_{\Ksf}(j_1)}, \mathbf{W}_{d_{k,2},\Nc_{\Ksf}(j_2)} \right)$ into $F_{k, \Nc_{\Ksf}(j_1) \cap \Nc_{\Ksf}(j_2)}$, which represents the set of coded symbols desired by user~$k$ that can be reconstructed by users in $\Nc_{\Ksf}(j_1) \cap \Nc_{\Ksf}(j_2)$.

\end{itemize}

The following lemma is proved in Appendix~\ref{sec: proof of second approach lemma}.
\begin{lem}
\label{lem: second approach}
For each $i \in [0: t_{\Ksf}+1]$ and  $k\in [\Ksf]$, we have 
\begin{align}
|F_{k, \Vc}|&= 
\left(  \left( \frac{\alpha_{\Ksf} \asf}{\binom{\Ksf}{t_{\Ksf}}} \right)^2 \binom{\Ksf-i}{t_{\Ksf}-i} \binom{\Ksf-t_{\Ksf}}{t_{\Ksf}-i }+ 
\left(\frac{(1- \alpha_{\Ksf}) \asf}{\binom{\Ksf}{t_{\Ksf}+1}}\right)^2  \binom{\Ksf-i}{t_{\Ksf}+1-i} \binom{\Ksf-t_{\Ksf}-1}{t_{\Ksf}+1-i } + 
\right. \nonumber \\&  \quad \left. 
+2   \frac{\alpha_{\Ksf}(1- 
\alpha_{\Ksf}) \asf^2}{\binom{\Ksf}{t_{\Ksf}}\binom{\Ksf}{t_{\Ksf}+1}}   \binom{\Ksf-i}{t_{\Ksf}-i} \binom{\Ksf-t_{\Ksf}}{t_{\Ksf}+1-i }\right) \ssf^2  ,  \label{eq:sec approach lem}
\end{align}
for all  $\Vc \subseteq ([\Ksf]\setminus \{k\})$ where $|\Vc|=i$.  
\end{lem}
In other words, the length of $F_{k,\Vc}$ only depends on $|\Vc|$. Hence, we define $f_{i,\asf}$ as the RHS of~\eqref{eq:sec approach lem}, representing the length of each $F_{k,\Vc}$ where $|\Vc|=i$.

The transmission is divided into $t_{\Ksf}+2$ rounds.  In round $i \in [0:t_{\Ksf}+1]$, for each $\Sc\subseteq [\Ksf]$ where $|\Sc|=i+1$,
the server broadcasts
\begin{align}
X_{\Sc}=\sum_{k \in \Sc} F_{k,\Sc\setminus \{k\}}, \label{eq:round i sec app}
\end{align} 
 such that each user~$k \in \Sc$ can recover $F_{k,\Sc\setminus \{k\}}$.
 
 Considering all  the $t_{\Ksf}+2$ rounds, the total load is 
 \begin{align}
 &  \frac{\sum_{i \in [0:t_{\Ksf}+1]} \binom{\Ksf}{i+1} f_{i,\asf} }{f(\rsf,\ssf,\rsf)}
 =  \frac{\sum_{i \in [0:t_{\Ksf}+1]} \binom{\Ksf}{i+1} f_{i,\asf} }{\asf^2 \ssf^2}, \label{eq:expand f for a<1}
 \end{align}
where~\eqref{eq:expand f for a<1} follows from $\asf\leq 1$. From~\eqref{eq:expand f for a<1}, we prove~\eqref{eq:R_col} for the case where $\asf \leq 1$.

 \subsubsection{$\asf > 1$}
\label{subsub:case a >1}
 
 We then consider the case where $\asf>1$ (i.e., $\rsf > \ssf$). In this case, each demanded matrix product is not full-rank. So compared to the proposed column-partition scheme for $\asf \leq 1$, we will use a novel coded cache placement and some
additional steps in the delivery phase to deal with the rank deficiency.
 \paragraph*{{ Placement phase}} 
 We partition each matrix $\mathbf{W}_{i}$ where $i\in [\Nsf]$ into two blocks 
\begin{align}
 (\mathbf{W}_{i})_{\ssf \times \rsf}  = \left[\begin{array}{c:c}
 (\mathbf{W}_{i,1})_{\ssf \times \ssf}  &    (\mathbf{W}_{i,2})_{\ssf \times (\rsf-\ssf)}  
\end{array}
\right].\label{eq:col block partition}
\end{align}
 Up to a column permutation,  the rank of $\mathbf{W}_{i,1}$ is equal to the rank of $\mathbf{W}_{i}$.\footnote{\label{foot:permutation} The information of permutation is also cached by each user, which is negligible  compared to the field size $\qsf$ and the cache size of each user.}

 The cache placement for $\mathbf{W}_{i,1}$ is the same as the case where $\asf <1$.  Recall that $ t_{\Ksf}=\left\lfloor \frac{ \Ksf \Msf}{\Nsf} \right\rfloor$ and $\alpha_{\Ksf}= t_{\Ksf}+1 -\frac{\Ksf \Msf}{\Nsf}$. The first $\alpha_{\Ksf} \ssf$ columns of $\mathbf{W}_{i,1}$   are partitioned into $\binom{\Ksf}{t_{\Ksf}}$ sub-matrices, each of which is denoted by $\mathbf{W}_{i,1, \Tc_1}$ and cached by users in $\Tc_1$ where $\Tc_1 \subseteq [\Ksf]$ and $|\Tc_1|=t_{\Ksf}$.  
 The remaining  $(1-\alpha_{\Ksf}) \ssf$ columns of $\mathbf{W}_{i,1}$ are partitioned into $\binom{\Ksf}{t_{\Ksf}+1}$ sub-matrices, each of which is denoted by $\mathbf{W}_{i,1,\Tc_2}$ and cached by users in $\Tc_2$ where $\Tc_2 \subseteq [\Ksf]$ and $|\Tc_2|=t_{\Ksf}+1$. 
 
 The cache placement for $\mathbf{W}_{i,2}$ is as follows. 
\begin{itemize}
\item We partition the first   $\alpha_{\Ksf} (\rsf-\ssf)$ columns of $\mathbf{W}_{i,2}$   into $\binom{\Ksf}{t_{\Ksf}}$ sub-matrices, each of which is denoted by $\mathbf{W}_{i,2,\Tc_1}$ where $\Tc_1 \subseteq [\Ksf]$ and $|\Tc_1|=t_{\Ksf}$.  $\mathbf{W}_{i,2,\Tc_1}$ has dimension $  \ssf \times \frac{\alpha_{\Ksf} (\rsf-\ssf)}{\binom{\Ksf}{t_{\Ksf}}}$.
  We let each user in $\Tc_1$ cache 
$
{\bf Q}(\mathbf{W}_{i,1},  \mathbf{W}_{i,2,\Tc_1}),
$
where 
$$\mathbf{W}_{i,1} {\bf Q}(\mathbf{W}_{i,1},  \mathbf{W}_{i,2,\Tc_1})=\mathbf{W}_{i,2,\Tc_1},$$ 
and
 the dimension of ${\bf Q}(\mathbf{W}_{i,1},  \mathbf{W}_{i,2,\Tc_1})$ is the same as $\mathbf{W}_{i,2,\Tc_1}$. 
More precisely, since  the rank of $\mathbf{W}_{i,1}$ is equal to the rank of $\mathbf{W}_{i}$, each column   of $\mathbf{W}_{i,2,\Tc_1}$ can be expressed by a linear combination  of the columns of $\mathbf{W}_{i,1}$. For example, the $j^{\text{th}}$  column   of $\mathbf{W}_{i,2,\Tc_1}$ is equal to 
$$
\mathbf{W}_{i,1} {\bf Q}_j(\mathbf{W}_{i,1},  \mathbf{W}_{i,2,\Tc_1}),
$$
where ${\bf Q}_j(\mathbf{W}_{i,1},  \mathbf{W}_{i,2,\Tc_1})$ represents the $j^{\text{th}}$  column of  $  {\bf Q}(\mathbf{W}_{i,1},  \mathbf{W}_{i,2,\Tc_1})$. Note that if $\mathbf{W}_{i,1}$ is full-rank, 
$
{\bf Q}(\mathbf{W}_{i,1},  \mathbf{W}_{i,2,\Tc_1}) $ becomes $ \mathbf{W}^{-1}_{i,1}  \mathbf{W}_{i,2,\Tc_1}.
$
\item   Similarly,
  the remaining  $(1-\alpha_{\Ksf})  (\rsf-\ssf)$ columns of $\mathbf{W}_{i,2}$ are partitioned into $\binom{\Ksf}{t_{\Ksf}+1}$ sub-matrices, each of which is denoted by $\mathbf{W}_{i,2,\Tc_2}$ where $\Tc_2 \subseteq [\Ksf]$ and $|\Tc_2|=t_{\Ksf}+1$. $\mathbf{W}_{i,2,\Tc_2}$ has dimension $  \ssf \times \frac{(1-\alpha_{\Ksf}) (\rsf-\ssf)}{\binom{\Ksf}{t_{\Ksf}+1}} $.
 We let each user in $\Tc_2$ cache 
$
  {\bf Q}(\mathbf{W}_{i,1}  \mathbf{W}_{i,2,\Tc_2}).
$
\end{itemize}

 Since the dimension of  $  {\bf Q}(\mathbf{W}_{i,1},  \mathbf{W}_{i,2,\Tc}) $ is the same as $\mathbf{W}_{i,2,\Tc}$ for any  $\Tc  \subseteq [\Ksf]$ where $|\Tc| \in \{t_{\Ksf}, t_{\Ksf}+1\}$,   the total number of symbols cached by each user is the same as for the case where $\asf \leq 1$ (which is $\Msf \ssf \rsf$). Hence the cache size constraint is satisfied.
 
  \paragraph*{{ Delivery phase}}
 The matrix product desired by user~$k\in [\Ksf]$ can be expressed as 
 \begin{subequations}
 \begin{align}
 (\mathbf{W}^{\text{\rm T}}_{d_{k,1}} \mathbf{W}_{d_{k,2}})_{\rsf \times \rsf}  
& 
 = \left[\begin{array}{c}
 (\mathbf{W}^{\text{\rm T}}_{d_{k,1},1})_{\ssf \times \ssf}   \\ \hdashline
 (\mathbf{W}^{\text{\rm T}}_{d_{k,1},2})_{(\rsf-\ssf) \times \ssf}  
\end{array}
\right] ~ \left[\begin{array}{c:c}
 (\mathbf{W}_{d_{k,2},1})_{\ssf \times \ssf}    &
 (\mathbf{W}_{d_{k,2},2})_{\ssf \times (\rsf-\ssf) }  
\end{array}
\right]  \\
&= \left[\begin{array}{c:c}
 (\mathbf{W}^{\text{\rm T}}_{d_{k,1},1} \mathbf{W}_{d_{k,2},1})_{\ssf \times \ssf}   &    (\mathbf{W}^{\text{\rm T}}_{d_{k,1},1} \mathbf{W}_{d_{k,2},2})_{\ssf \times (\rsf-\ssf)} \\ \hdashline
  (\mathbf{W}^{\text{\rm T}}_{d_{k,1},2} \mathbf{W}_{d_{k,2},1})_{(\rsf-\ssf) \times \ssf}   & (\mathbf{W}^{\text{\rm T}}_{d_{k,1},2} \mathbf{W}_{d_{k,2},2})_{ (\rsf-\ssf) \times (\rsf-\ssf) }  
\end{array}
\right].
 \label{eq:col demand block partition}
\end{align}
\end{subequations}
In the following, we divide the transmission into three steps.
 
First step: we deliver packets for  $\mathbf{W}^{\text{\rm T}}_{d_{k,1},1} \mathbf{W}_{d_{k,2},1}$, for all $k\in [\Ksf]$. 
The transmission for   $\mathbf{W}^{\text{\rm T}}_{d_{k,1},1} \mathbf{W}_{d_{k,2},1}$ is the same as the proposed column-partition scheme with $\asf=1$ as described earlier in this subsection. 
  Thus with the same derivation that led to~\eqref{eq:expand f for a<1}, the total number of symbols transmitted in the first step is 
  \begin{align}
 \sum_{i \in [0:t_{\Ksf}+1]} \binom{\Ksf}{i+1} f_{i,1} =y \ssf^2, \label{eq:load for block 1}
  \end{align}
 where $y$ is defined in~\eqref{eq:def of y}.
 
Second step: we then focus on   $\mathbf{W}^{\text{\rm T}}_{d_{k,1},1} \mathbf{W}_{d_{k,2},2}$. We partition $\mathbf{W}^{\text{\rm T}}_{d_{k,1},1} \mathbf{W}_{d_{k,2},2}$ into $\binom{\Ksf}{t_{\Ksf}}+\binom{\Ksf}{t_{\Ksf}+1}$ sub-matrices as 
 \begin{align}
 \mathbf{W}^{\text{\rm T}}_{d_{k,1},1} \mathbf{W}_{d_{k,2},2}=
 \left[\begin{array}{c:c:c}
 \mathbf{W}^{\text{\rm T}}_{d_{k,1},1}  \mathbf{W}_{d_{k,2},2, \Nc_{\Ksf} (1) }   & \cdots &
 \mathbf{W}^{\text{\rm T}}_{d_{k,1},1}  \mathbf{W}_{d_{k,2},2, \Nc_{\Ksf} \left(\binom{\Ksf+1}{t_{\Ksf}+1} \right) }  
\end{array}
\right].
\end{align}
For each $j \in \left[\binom{\Ksf+1}{t_{\Ksf}+1} \right]$, we have 
\begin{align}
 \mathbf{W}^{\text{\rm T}}_{d_{k,1},1}   \mathbf{W}_{d_{k,2},2, \Nc_{\Ksf} (j) }  
=  \mathbf{W}^{\text{\rm T}}_{d_{k,1},1}   \mathbf{W}_{d_{k,2},1}     {\bf Q}\left( \mathbf{W}_{d_{k,2},1} ,  \mathbf{W}_{d_{k,2},2, \Nc_{\Ksf} (j) } \right) .
  \label{eq:seonc block partition}
\end{align}
 Note that $\mathbf{W}^{\text{\rm T}}_{d_{k,1},1} \mathbf{W}_{d_{k,2},1} $ has been recovered by user~$k$ in the first delivery step. Hence, in this step user~$k$ needs to recover $  {\bf Q}( \mathbf{W}_{d_{k,2},1}   \mathbf{W}_{d_{k,2},2, \Nc_{\Ksf} (j) })$, which is cached by users in $\Nc_{\Ksf}(j)$. 
 We let 
 $$
 F^{\prime}_{k,\Nc_{\Ksf} (j) }=    {\bf Q}(\mathbf{W}_{d_{k,2},1},  \mathbf{W}_{d_{k,2},2, \Nc_{\Ksf} (j) }),
 $$
 which contains $ \ssf   \frac{\alpha_{\Ksf} (\rsf-\ssf)}{\binom{\Ksf}{t_{\Ksf}}}$ symbols if $|\Nc_{\Ksf}(j)|=t_{\Ksf}$, and contains $  \ssf  \frac{(1-\alpha_{\Ksf}) (\rsf-\ssf)}{\binom{\Ksf}{t_{\Ksf}+1}} $ symbols if $|\Nc_{\Ksf}(j)|=t_{\Ksf}+1$.
 
For each set $\Sc_1 \subseteq [\Ksf]$ where $\Sc_1=t_{\Ksf}+1$, the server broadcasts
\begin{align}
  \sum_{j \in \Sc_1}  F^{\prime}_{k,\Sc_1\setminus \{k\} } .  
\end{align} 
For each set   $\Sc_2 \subseteq [\Ksf]$ where $\Sc_2=t_{\Ksf}+2$, the server broadcasts
\begin{align}
  \sum_{j \in \Sc_2} F^{\prime}_{k,\Sc_2\setminus \{k\} } . 
\end{align} 
Hence, the total number of symbols transmitted in the second step is 
 \begin{align}
 & \binom{\Ksf}{t_{\Ksf}+1} \ssf   \frac{\alpha_{\Ksf} (\rsf-\ssf)}{\binom{\Ksf}{t_{\Ksf}}}+ \binom{\Ksf}{t_{\Ksf}+2}  \ssf  \frac{(1-\alpha_{\Ksf}) (\rsf-\ssf)}{\binom{\Ksf}{t_{\Ksf}+1}}.\label{eq:load for block 2}
 \end{align}

Third step: we let each user~$k \in [\Ksf]$ recover the remaining parts of its desired matrix product
 \begin{subequations}
 \begin{align}
 &\left[\begin{array}{c:c}
   \mathbf{W}^{\text{\rm T}}_{d_{k,1},2} \mathbf{W}_{d_{k,2},1}   &  \mathbf{W}^{\text{\rm T}}_{d_{k,1},2} \mathbf{W}_{d_{k,2},2} 
\end{array}
\right]  
\nonumber\\
&=\left[\begin{array}{c:c}
\left( \mathbf{W}_{d_{k,1},1}  {\bf Q}(\mathbf{W} _{d_{k,1},1} ,   \mathbf{W}_{d_{k,1},2})   \right) ^{\text{\rm T}}     \mathbf{W}_{d_{k,2},1}&
\left(  \mathbf{W}_{d_{k,1},1}  {\bf Q}(\mathbf{W}_{d_{k,1},1}  ,  \mathbf{W}_{d_{k,1},2} )   \right) ^{\text{\rm T}}   \mathbf{W}_{d_{k,2},2}
\end{array}
\right]\\
&=
 \left[\begin{array}{c:c}
\left(    {\bf Q}(\mathbf{W} _{d_{k,1},1} ,   \mathbf{W}_{d_{k,1},2})   \right) ^{\text{\rm T}}    \mathbf{W}^{\text{\rm T}}_{d_{k,1},1} \mathbf{W}_{d_{k,2},1}&
\left(   {\bf Q}(\mathbf{W}_{d_{k,1},1}  ,  \mathbf{W}_{d_{k,1},2} )   \right) ^{\text{\rm T}}  \mathbf{W}^{\text{\rm T}}_{d_{k,1},1} \mathbf{W}_{d_{k,2},2}
\end{array}
\right].
 \label{eq:col 3 4 block partition}
\end{align}
\end{subequations}
Note that  $ \mathbf{W}^{\text{\rm T}}_{d_{k,1},1} \mathbf{W}_{d_{k,2},1}$ and $\mathbf{W}^{\text{\rm T}}_{d_{k,1},1} \mathbf{W}_{d_{k,2},2}$ have been recovered by user~$k$ in the first and second steps, respectively. Now it only needs to recover  ${\bf Q}(\mathbf{W}_{d_{k,1},1}    \mathbf{W}_{d_{k,1},2})$, which can be expressed as
\begin{align}
{\bf Q}(\mathbf{W}_{d_{k,1},1}  ,  \mathbf{W}_{d_{k,1},2})=
 \left[\begin{array}{c:c:c}
{\bf Q}\left(\mathbf{W}_{d_{k,1},1} , \mathbf{W}_{d_{k,1},2, \Nc_{\Ksf} (1) } \right)  & \cdots &
{\bf Q}\left(\mathbf{W}_{d_{k,1},1} , \mathbf{W}_{d_{k,1},2, \Nc_{\Ksf} \left(\binom{\Ksf+1}{t_{\Ksf}+1} \right) }  \right)
\end{array}
\right] 
\end{align}
For each $j \in \left[\binom{\Ksf+1}{t_{\Ksf}+1} \right]$,
 we let 
 $$
 F^{\prime\prime}_{k,\Nc_{\Ksf} (j) }= {\bf Q}\left( \mathbf{W}_{d_{k,1},1} , \mathbf{W}_{d_{k,1},2, \Nc_{\Ksf} (j) } \right),
 $$
 which contains $ \ssf   \frac{\alpha_{\Ksf} (\rsf-\ssf)}{\binom{\Ksf}{t_{\Ksf}}}$ symbols if $|\Nc_{\Ksf}(j)|=t_{\Ksf}$, and contains $  \ssf  \frac{(1-\alpha_{\Ksf}) (\rsf-\ssf)}{\binom{\Ksf}{t_{\Ksf}+1}} $ symbols if $|\Nc_{\Ksf}(j)|=t_{\Ksf}+1$.
 
For each set $\Sc_1 \subseteq [\Ksf]$ where $\Sc_1=t_{\Ksf}+1$, the server broadcasts
\begin{align}
  \sum_{j \in \Sc_1}  F^{\prime\prime}_{k,\Sc_1\setminus \{k\} } .  
\end{align} 
For each set   $\Sc_2 \subseteq [\Ksf]$ where $\Sc_2=t_{\Ksf}+2$, the server broadcasts
\begin{align}
  \sum_{j \in \Sc_2} F^{\prime\prime}_{k,\Sc_2\setminus \{k\} } . 
\end{align} 
Hence, the total number of symbols transmitted in the third step is 
 \begin{align}
 & \binom{\Ksf}{t_{\Ksf}+1} \ssf   \frac{\alpha_{\Ksf} (\rsf-\ssf)}{\binom{\Ksf}{t_{\Ksf}}}+ \binom{\Ksf}{t_{\Ksf}+2}  \ssf  \frac{(1-\alpha_{\Ksf}) (\rsf-\ssf)}{\binom{\Ksf}{t_{\Ksf}+1}}.\label{eq:load for block 3}
 \end{align}

Considering all the three steps, from~\eqref{eq:load for block 1},~\eqref{eq:load for block 2}, and~\eqref{eq:load for block 3}, the total load is 
\begin{align}
\frac{y \ssf^2+ 2 \binom{\Ksf}{t_{\Ksf}+1} \ssf   \frac{\alpha_{\Ksf} (\rsf-\ssf)}{\binom{\Ksf}{t_{\Ksf}}}+ 2 \binom{\Ksf}{t_{\Ksf}+2}  \ssf  \frac{(1-\alpha_{\Ksf}) (\rsf-\ssf)}{\binom{\Ksf}{t_{\Ksf}+1}}}{f(\rsf,\ssf,\rsf)}= \frac{y + 2 \binom{\Ksf}{t_{\Ksf}+1}   \frac{\alpha_{\Ksf} (\asf-1)}{\binom{\Ksf}{t_{\Ksf}}}+ 2 \binom{\Ksf}{t_{\Ksf}+2}  \frac{(1-\alpha_{\Ksf}) (\asf-1)}{\binom{\Ksf}{t_{\Ksf}+1}}}{ 2\asf-1 }, \label{eq:load col a>1} 
\end{align} 
where~\eqref{eq:load col a>1} follows from that $\asf>1$.
From~\eqref{eq:load col a>1}, we prove~\eqref{eq:R_col} for the case where $\asf>1$.
 
\begin{rem}[Column-partition scheme v.s. uncoded caching baseline scheme]
 \label{rem:proof cor formula 2}
 \rm
 When $\asf \leq 1$, for any pair $(j_1,j_2)$ where $j_1, j_2 \in \left[ \binom{\Ksf+1}{t_{\Ksf}+1}\right]$ and $k \notin \Nc_{\Ksf}(j_1) \cap \Nc_{\Ksf}(j_2)$, if the server directly broadcasts $\mathbf{W}^{\text{\rm T}}_{d_{k,1},\Nc_{\Ksf}(j_1)} \mathbf{W}_{d_{k,2},\Nc_{\Ksf}(j_2)} $, then our column-partition scheme reduces to the uncoded caching baseline scheme in Theorem~\ref{thm:base}. Hence, in this case our column-partition scheme is strictly better than the uncoded caching baseline scheme if $0<\Msf<\Nsf$. When $\asf>1$, besides the above improvement which also appears in the first delivery step, in the second and third steps we further compress the desired matrix products of  the users by leveraging the correlation among the elements in the product and the users' caches. Hence, in this case our column-partition scheme is strictly better than the uncoded caching baseline scheme if $ \Msf<\Nsf$.
  \hfill $\square$ 
 \end{rem} 

 \section{Conclusions}
\label{sec:conclusion} 
This paper introduced a novel coded caching problem for matrix multiplication retrieval, where each cache-aided user requests the product of two matrices in the library. We first proposed 
a structure-agnostic scheme which treats each product as an independent file. In order to leverage the structure of matrix multiplications, 
we proposed two schemes (row-partition and column-partition) to attain coded caching gain for the matrix multiplication retrieval problem, by leveraging the correlation among the elements in each product. The proposed schemes outperform the baseline schemes. 
For ``fat'' matrices, 
the proposed row-partition scheme is proved to be order optimal   within a factor of $2$ under the constraint of uncoded cache placement and  $\Nsf \geq 2\Ksf$.  


\appendices

   \section{Structure-agnostic Scheme: Proof of Theorem~\ref{thm:load of agnostic}}
\label{sub:agnostic}
For each pair $(i,j)$ where $1\leq i\leq  j \leq \Nsf$, we define  
\begin{align}
W_{(i,j)}:= P\left(\mathbf{W}^{\text{\rm T}}_i, \mathbf{W}_j  \right) 
\end{align}
and treat $W_{(i,j)}$
as an independent file with $\Bsf$ symbols, where we can recover $\mathbf{W}^{\text{\rm T}}_i  \mathbf{W}_j $  from $W_{(i,j)}$.
We 
then use the   MAN coded caching scheme as follows.

 \paragraph*{{ Placement phase}}
 We focus on each $t\in [0:\Ksf]$.
 For each pair $(i,j)$ where $1\leq i\leq  j \leq \Nsf$, we divide $W_{(i,j)}$ into $\binom{\Ksf}{t}$ non-overlapping and equal-length subfiles, $W_{(i,j)}=\{W_{(i,j),\Tc}:\Tc \subseteq [\Ksf], |\Tc|=t\}$, where each subfile $W_{(i,j),\Tc}$ contains $\frac{\Bsf}{\binom{\Ksf}{t}}$ symbols and  is cached by users in $\Tc$.
 As  there are $\binom{\Nsf}{2}+\Nsf= \frac{\Nsf(\Nsf+1)}{2}$ pairs $(i,j)$ where $1\leq i\leq  j \leq \Nsf$, 
 the total number of symbols cached by each user is 
 $$
\frac{\Nsf(\Nsf+1)}{2} \frac{\Bsf \binom{\Ksf-1}{t-1} }{\binom{\Ksf}{t}}= \frac{\Nsf(\Nsf+1)\Bsf  t}{2\Ksf} = \frac{\Nsf(\Nsf+1)\Bsf  t}{2\asf\Ksf \ssf^2}\ssf \rsf=\Msf \ssf \rsf,
 $$
satisfying the cache size constraint.

 \paragraph*{{ Delivery phase}}
Each user  $k \in [\Ksf]$ demands $W_{\dv_k}$. For each set $\Sc \subseteq [\Ksf]$ where $|\Sc|=t+1$, the server transmits
\begin{align}
\sum_{k \in \Sc} W_{\dv_k,\Sc \setminus \{k\}},   \label{eq:structure agnostic multicast} 
\end{align}
where each user~$k\in \Sc$ caches all subfiles except $ W_{\dv_k,\Sc \setminus \{k\}}$ such that it can recover $ W_{\dv_k,\Sc \setminus \{k\}}$.

After considering all sets of users with cardinality $t+1$, each user can recover its demanded file and thus recover its demanded product. Hence, the total load is  
$$
\frac{ \binom{\Ksf}{t+1} }{\binom{\Ksf}{t}}= \frac{\Ksf-t}{t+1},
$$
 which coincides  with~\eqref{eq:load of agnostic}.
 
\section{Proof of Theorem~\ref{thm:order optimality}}
\label{sec:converse proof}   
We consider the case where $\asf \geq 1$ and $\Nsf \geq 2\Ksf$.  
\subsection{Converse}   
   \label{sub:theorem converse}
We consider the worse-case demands, where $[\dv_1;\ldots;\dv_{\Ksf}]$ contains $2 \Ksf$ different indices of matrices.    
   
We use a genie-aided converse bound. We assume that during the delivery phase there is a private link from the server to each user~$k \in [\Ksf]$ through which  the server transmits $\mathbf{W}^{\text{\rm T}}_{d_{k,1}}$ to user~$k$. In this case, 
the   minimum worst-case number of broadcasted symbols by the server under uncoded cache placement   is denoted by $\Lsf^{\star}_{\text{genie, u}}$. 
Obviously, we have 
\begin{align}
&\Rsf^{\star}_{\text{u}} f(\rsf,\ssf, \rsf) \geq \Lsf^{\star}_{\text{genie, u}} .\label{eq:genie 1 uncoded} 
\end{align}
 Recall that $\mathbf{W}^{\text{\rm T}}_{d_{k,1}}$  is  of  dimension $\rsf \times \ssf$ where $\rsf \geq \ssf$ and its elements are uniformly i.i.d. Hence, if 
    user~$k$ can recover  $\mathbf{W}^{\text{\rm T}}_{d_{k,1}} \mathbf{W}_{d_{k,2}}$, with the knowledge of $\mathbf{W}^{\text{\rm T}}_{d_{k,1}}$  this user can also recover $\mathbf{W}_{d_{k,2}}$. 
    On the other hand, if  user~$k$ can  recover $\mathbf{W}_{d_{k,2}}$, with the knowledge of $\mathbf{W}^{\text{\rm T}}_{d_{k,1}}$  this user can also recover  $\mathbf{W}^{\text{\rm T}}_{d_{k,1}} \mathbf{W}_{d_{k,2}}$. Hence, when $\asf \geq 1$ we have 
    $$H(\mathbf{W}^{\text{\rm T}}_{d_{k,1}}\mathbf{W}_{d_{k,2}}| \mathbf{W}^{\text{\rm T}}_{d_{k,1}})=H( \mathbf{W}_{d_{k,2}}| \mathbf{W}^{\text{\rm T}}_{d_{k,1}})=H( \mathbf{W}_{d_{k,2}} ).$$ 
    Under the constraint of uncoded cache placement, it is equivalent to the problem
      with the same network but each user aims to retrieve a whole file (each file has $\ssf \rsf$ symbols). 

In addition, since the cache placement is uncoded and   $[\dv_1;\ldots;\dv_{\Ksf}]$ contains $2 \Ksf$ different indices of matrices, the matrix transmitted through the private link cannot help each user~$k \in [\Ksf]$ 
to decode its desired file (i.e., $\mathbf{W}_{d_{k,2}}$). Thus we can use the converse bound in~\cite{indexcodingcaching2020,exactrateuncoded} for the original MAN coded caching problem for  single file  retrieval  to lower bound $\Lsf^{\star}_{\text{genie, u}}$. In other words, 
  $(\Msf, \Lsf^{\star}_{\text{genie, u}})$ is lower bounded by the lower convex envelop of $\left(\frac{\Nsf t}{\Ksf},  \frac{\Ksf-t}{t+1}\ssf\rsf/ \Bsf \right)$, for all $t\in [0:\Ksf]$.
 In conclusion, from~\eqref{eq:genie 1 uncoded}, 
  $(\Msf, \Rsf^{\star}_{\text{u}} )$ is   lower bounded by the lower convex envelop of 
\begin{align}
\left(\frac{\Nsf t}{\Ksf},  \frac{\Ksf-t}{t+1} \frac{ \ssf\rsf }{f(\rsf,\ssf, \rsf)}\right)=\left(\frac{\Nsf t}{\Ksf},  \frac{\Ksf-t}{t+1} \frac{ \asf }{2\asf -1 }\right), \ \forall t\in [0:\Ksf]. \label{eq:final converse}
\end{align}
  
\subsection{Achievability}   
   \label{sub:theorem achie}
   From~\eqref{eq:second baseline},  the multi-request baseline scheme can achieve the lower convex envelop of $(\Msf, \Rsf_2)= \left(\frac{\Nsf t}{\Ksf},  \frac{2(\Ksf-t) \asf }{(t+1)g(\asf,\asf)} \right)$, for all $t\in [0:\Ksf]$.
   Compared with the converse bound in~\eqref{eq:final converse}, the multi-request baseline scheme is order optimal   within a factor of $2$ under the constraint of uncoded cache placement and $\asf \geq 1$. In addition, from Corollary~\ref{cor:improve},  the proposed row-partition scheme outperforms  the multi-request baseline scheme. Hence, we prove Theorem~\ref{thm:order optimality}.
   
\section{Proof of Lemma~\ref{lem: second approach}}
\label{sec: proof of second approach lemma}
We fix one $k\in [\Ksf]$ and one $i \in [0: t_{\Ksf}+1]$. Now we want to compute the length of $F_{k,\Vc}$ where $\Vc\subseteq ([\Ksf]\setminus \{k\})$ and $|\Vc|=i$.
If one  pair $(j_1, j_2)$ where  $j_1, j_2 \in \left[ \binom{\Ksf+1}{t_{\Ksf}+1}\right]$ satisfies that, $\Nc_{\Ksf}(j_1)\cap \Nc_{\Ksf}(j_2)= \Vc$, we have that $F_{k,\Vc}$ contains $P\left( \mathbf{W}^{\text{\rm T}}_{d_{k,1},\Nc_{\Ksf}(j_1)}, \mathbf{W}_{d_{k,2},\Nc_{\Ksf}(j_2)} \right)$.

Now we divide all the pairs  $(j_1, j_2)$ where $j_1, j_2 \in \left[ \binom{\Ksf+1}{t_{\Ksf}+1}\right]$ and $\Nc_{\Ksf}(j_1)\cap \Nc_{\Ksf}(j_2)= \Vc$ into the following four cases.


\paragraph*{{ Case 1: $|\Nc_{\Ksf}(j_1)|=|\Nc_{\Ksf}(j_2)|=t_{\Ksf}$}}
In this case,   the length of $P\left( \mathbf{W}^{\text{\rm T}}_{d_{k,1},\Nc_{\Ksf}(j_1)}, \mathbf{W}_{d_{k,2},\Nc_{\Ksf}(j_2)} \right)$    is 
 \begin{subequations}
\begin{align}
f\left(\frac{\alpha_{\Ksf} \rsf}{\binom{\Ksf}{t_{\Ksf}}},  \ssf, \frac{\alpha_{\Ksf} \rsf}{\binom{\Ksf}{t_{\Ksf}}}  \right)&=  g \left(\frac{\alpha_{\Ksf} \asf}{\binom{\Ksf}{t_{\Ksf}}},\frac{\alpha_{\Ksf} \asf}{\binom{\Ksf}{t_{\Ksf}}} \right) \ssf^2   \\
&=\left(\frac{\alpha_{\Ksf} \asf}{\binom{\Ksf}{t_{\Ksf}}}\right)^2 \ssf^2, \label{eq:because a<=1}
\end{align}
 \end{subequations}
 where~\eqref{eq:because a<=1} comes from that $\asf \leq 1$ and thus $\frac{\alpha_{\Ksf} \asf}{\binom{\Ksf}{t_{\Ksf}}}\leq 1$.
The number of pairs  $(j_1, j_2)$  where $j_1, j_2 \in \left[ \binom{\Ksf+1}{t_{\Ksf}+1}\right]$, $\Nc_{\Ksf}(j_1)\cap \Nc_{\Ksf}(j_2)= \Vc$, and $|\Nc_{\Ksf}(j_1)|=|\Nc_{\Ksf}(j_2)|=t_{\Ksf}$ is 
\begin{align}
\binom{\Ksf-|\Vc|}{t_{\Ksf}-|\Vc|} \binom{\Ksf-t_{\Ksf}}{t_{\Ksf}-|\Vc|}=\binom{\Ksf-i}{t_{\Ksf}-i} \binom{\Ksf-t_{\Ksf}}{t_{\Ksf}-i}.
\end{align}

\paragraph*{{ Case 2: $|\Nc_{\Ksf}(j_1)|=|\Nc_{\Ksf}(j_2)|=t_{\Ksf}+1$}}
In this case,   the length of $P\left( \mathbf{W}^{\text{\rm T}}_{d_{k,1},\Nc_{\Ksf}(j_1)}, \mathbf{W}_{d_{k,2},\Nc_{\Ksf}(j_2)} \right)$    is 
 \begin{subequations}
\begin{align}
f\left(\frac{(1-\alpha_{\Ksf}) \rsf}{\binom{\Ksf}{t_{\Ksf}+1}},  \ssf, \frac{(1-\alpha_{\Ksf}) \rsf}{\binom{\Ksf}{t_{\Ksf}+1}} \right)&=  g \left(\frac{(1-\alpha_{\Ksf}) \asf}{\binom{\Ksf}{t_{\Ksf}+1}},\frac{(1-\alpha_{\Ksf}) \asf}{\binom{\Ksf}{t_{\Ksf}+1}} \right) \ssf^2   \\
&= \left( \frac{(1-\alpha_{\Ksf}) \asf}{\binom{\Ksf}{t_{\Ksf}+1}}\right)^2 \ssf^2.
\end{align}
 \end{subequations}
The number of pairs  $(j_1, j_2)$  where $j_1, j_2 \in \left[ \binom{\Ksf+1}{t_{\Ksf}+1}\right]$, $\Nc_{\Ksf}(j_1)\cap \Nc_{\Ksf}(j_2)= \Vc$, and $|\Nc_{\Ksf}(j_1)|=|\Nc_{\Ksf}(j_2)|=t_{\Ksf}+1$ is 
\begin{align}
\binom{\Ksf-|\Vc|}{t_{\Ksf}+1-|\Vc|} \binom{\Ksf-t_{\Ksf}-1}{t_{\Ksf}+1-|\Vc|}=\binom{\Ksf-i}{t_{\Ksf}+1-i} \binom{\Ksf-t_{\Ksf}-1}{t_{\Ksf}+1-i}.
\end{align}

\paragraph*{{ Case 3: $|\Nc_{\Ksf}(j_1)|=t_{\Ksf}$ and $|\Nc_{\Ksf}(j_2)|=t_{\Ksf}+1$}}
In this case,  the length of $P\left( \mathbf{W}^{\text{\rm T}}_{d_{k,1},\Nc_{\Ksf}(j_1)}, \mathbf{W}_{d_{k,2},\Nc_{\Ksf}(j_2)} \right)$    is 
 \begin{subequations}
\begin{align}
f\left(\frac{\alpha_{\Ksf} \rsf}{\binom{\Ksf}{t_{\Ksf}}},  \ssf, \frac{(1-\alpha_{\Ksf}) \rsf}{\binom{\Ksf}{t_{\Ksf}+1}}  \right)&=  g \left(\frac{\alpha_{\Ksf} \asf}{\binom{\Ksf}{t_{\Ksf}}},\frac{(1-\alpha_{\Ksf}) \asf}{\binom{\Ksf}{t_{\Ksf}+1}} \right) \ssf^2 \\
&=  \frac{\alpha_{\Ksf} (1-\alpha_{\Ksf})  \asf^2}{\binom{\Ksf}{t_{\Ksf}}\binom{\Ksf}{t_{\Ksf}+1}} \ssf^2.
\end{align}
 \end{subequations}
The number of pairs  $(j_1, j_2)$  where $j_1, j_2 \in \left[ \binom{\Ksf+1}{t_{\Ksf}+1}\right]$, $\Nc_{\Ksf}(j_1)\cap \Nc_{\Ksf}(j_2)= \Vc$,  $|\Nc_{\Ksf}(j_1)|=t_{\Ksf}$, and $|\Nc_{\Ksf}(j_2)|=t_{\Ksf}+1$ is 
\begin{align}
\binom{\Ksf-|\Vc|}{t_{\Ksf}-|\Vc|} \binom{\Ksf-t_{\Ksf}}{t_{\Ksf}+1-|\Vc|}=\binom{\Ksf-i}{t_{\Ksf}-i} \binom{\Ksf-t_{\Ksf}}{t_{\Ksf}+1-i}.
\end{align}

\paragraph*{{ Case 4: $|\Nc_{\Ksf}(j_1)|=t_{\Ksf}+1$ and $|\Nc_{\Ksf}(j_2)|=t_{\Ksf}$}}
In this case,    the length of $P\left( \mathbf{W}^{\text{\rm T}}_{d_{k,1},\Nc_{\Ksf}(j_1)}, \mathbf{W}_{d_{k,2},\Nc_{\Ksf}(j_2)} \right)$    is 
 \begin{subequations}
\begin{align}
f\left(\frac{(1-\alpha_{\Ksf}) \rsf}{\binom{\Ksf}{t_{\Ksf}+1}},  \ssf,\frac{\alpha_{\Ksf} \rsf}{\binom{\Ksf}{t_{\Ksf}}}   \right)&=  g \left( \frac{(1-\alpha_{\Ksf}) \asf}{\binom{\Ksf}{t_{\Ksf}+1}},\frac{\alpha_{\Ksf} \asf}{\binom{\Ksf}{t_{\Ksf}}} \right) \ssf^2 \\
&=  \frac{\alpha_{\Ksf} (1-\alpha_{\Ksf})  \asf^2}{\binom{\Ksf}{t_{\Ksf}}\binom{\Ksf}{t_{\Ksf}+1}} \ssf^2.
\end{align}
 \end{subequations}
The number of pairs  $(j_1, j_2)$  where $j_1, j_2 \in \left[ \binom{\Ksf+1}{t_{\Ksf}+1}\right]$, $\Nc_{\Ksf}(j_1)\cap \Nc_{\Ksf}(j_2)= \Vc$,  $|\Nc_{\Ksf}(j_1)|=t_{\Ksf}+1$, and $|\Nc_{\Ksf}(j_2)|=t_{\Ksf}$ is 
 \begin{subequations}
\begin{align}
\binom{\Ksf-|\Vc|}{t_{\Ksf}+1-|\Vc|} \binom{\Ksf-t_{\Ksf}-1}{t_{\Ksf}-|\Vc|} 
&=\binom{\Ksf-i}{t_{\Ksf}+1-i} \binom{\Ksf-t_{\Ksf}-1}{t_{\Ksf}-i} \\
&=\binom{\Ksf-i}{t_{\Ksf}-i} \binom{\Ksf-t_{\Ksf}}{t_{\Ksf}+1-i}.
\end{align}
 \end{subequations}

Considering all the above four cases, we can prove Lemma~\ref{lem: second approach}.
\bibliographystyle{IEEEtran}
\bibliography{IEEEabrv,IEEEexample}

\end{document}

%% file: macros.tex
\long\def\comment#1{}

\newcommand{\dv}{{\mathbf d}}

\newcommand{\Gc}{{\mathcal G}}

\newcommand{\Nc}{{\mathcal N}}

\newcommand{\Qc}{{\mathcal Q}}

\newcommand{\Sc}{{\mathcal S}}
\newcommand{\Tc}{{\mathcal T}}

\newcommand{\Vc}{{\mathcal V}}

\newcommand{\asf}{{\mathsf a}}

\newcommand{\qsf}{{\mathsf q}}
\newcommand{\rsf}{{\mathsf r}}
\newcommand{\ssf}{{\mathsf s}}

\newcommand{\Bsf}{{\mathsf B}}

\newcommand{\Ksf}{{\mathsf K}}
\newcommand{\Lsf}{{\mathsf L}}
\newcommand{\Msf}{{\mathsf M}}
\newcommand{\Nsf}{{\mathsf N}}

\newcommand{\Rsf}{{\mathsf R}}

\newtheorem{thm}{Theorem}
\newtheorem{cor}{Corollary}
\newtheorem{lem}{Lemma}

\newtheorem{rem}{Remark}
\newtheorem{example}{Example}

\providecommand{\definitionname}{Definition}